\documentclass[showpacs, prd, nofootinbib, twocolumn, floatfix, preprintnumbers]{revtex4}

\usepackage{amsmath}
\usepackage{amssymb}
\usepackage{color}
\usepackage{verbatim}
\usepackage{graphicx}
\usepackage[vcentermath]{youngtab}
\usepackage{amsthm}

\newcommand{\dif}{\mathrm{d}}

\newcommand{\mt}{\mathcal{T}}
\newcommand{\mvac}{M_\text{vac}}
\newcommand{\ai}{\ensuremath{a_1}}
\newcommand{\aii}{\ensuremath{a_2}}
\newcommand{\aiii}{\ensuremath{a_3}}
\newcommand{\tr}{\tilde{r}}
\newcommand{\tg}{\tilde{g}}

\DeclareMathOperator{\sech}{sech}

\theoremstyle{plain}
\newtheorem{theorem}{Theorem}
\newtheorem*{lemma}{Lemma}
\newtheorem*{conj}{Conjecture}

\newlength{\figwidth}
\begin{document}

\title{Dynamical Lorentz symmetry breaking and topological defects}

\author{Michael D.\ Seifert}

\affiliation{Dept.\ of Physics, Indiana University, 727 E.\
$\text{3}^\text{rd}$ St., Bloomington, IN, 47405}

\email{mdseifer@indiana.edu}

\begin{abstract}
  I discuss the possibility of topological defect solutions in field
  theories containing a tensor field which spontaneously breaks
  Lorentz symmetry.  I find that for theories of a tensor with rank $r
  \leq 5$ and for which the vacuum manifold consists of the tensors
  whose ``square'' is some constant value, only three types of tensor
  (vectors, antisymmetric two-tensors, and symmetric two-tensors) have
  the appropriate vacuum manifold topology to support topological
  defects.  Of these, topological defect solutions can be easily
  constructed for two: vector domain wall solutions and antisymmetric
  tensor monopole solutions.  These antisymmetric tensor monopole
  solutions are in principle detectible via their gravitational
  lensing effects.
\end{abstract}

\preprint{IUHET 547, August 2010}

\pacs{11.27.+d,11.30.Cp,11.30.Qc,14.80.-j}

\maketitle

\section{Introduction}

The idea of Lorentz symmetry violation has been a subject of sustained
research activity for some years now.  In such theories, one typically
postulates the existence of a non-zero tensor on spacetime that
couples to ``conventional'' matter such as electrons, quarks, photons,
and so forth.  The so-called ``Standard Model Extension'', or SME
\cite{CKSME}, provides a wide-ranging framework within which to
analyse the physical effects of such symmetry violations.  While no
unambiguous evidence for such fields has yet been found, many
experimental bounds on their effects have been obtained \cite{LVdata},
and research is ongoing.

It is, of course, natural to ask what the origin of this
``Lorentz-violating'' tensor might be.  It is known that the fiat
specification of a fixed background tensor field, while acceptable in
a flat spacetime, is in general not mathematically consistent with
making the metric dynamical \cite{Kosgrav}.  However, a
self-consistent theory can be obtained by allowing the
Lorentz-violating field to itself be dynamical.  In this scenario, the
tensor field is usually taken to have a potential energy that is
minimized (and vanishes) when the tensor takes on a non-zero value;
this non-zero value can be said to spontaneously break the Lorentz
symmetry of the underlying Lagrangian.  This scenario closely
parallels the behaviour of the Higgs field in the Standard Model;
however, in our case the field taking on a background expectation
value is a spacetime tensor rather than a spacetime scalar.  The usual
``particle physics'' portion of the SME, with conventional matter
fields coupling to a constant background tensor in flat spacetime, can
be obtained as an effective field theory limit with the dynamics of
the Lorentz-violating tensor field integrated out.

While this picture is self-consistent and fairly compelling, it also
raises an interesting corollary possibility.  In general, a field
which spontaneously breaks some symmetry \emph{in vacuo} will, at
sufficiently high temperatures, see that symmetry restored.  This
implies that in the early Universe, a Lorentz-violating tensor field
will have zero expectation value; as the Universe expands and cools,
we would eventually expect this field to undergo a phase transition
from a state of higher symmetry (a vanishing tensor field) to a state
of lower symmetry (a non-vanishing tensor field.)  

A field whose Lagrangian possesses some symmetry but whose solutions
break that symmetry will generally not have a unique minimum to its
potential; this space of all possible vacuum values is known as the
\emph{vacuum manifold}.  Since there is more than one possible vacuum
value for the field, it is likely that causally disconnected portions
of the Universe would ``choose'' different values of the field in the
symmetry-breaking phase transition.  Moreover, if this manifold has
particular topological properties, regions that ``fall into''
different portions of the vacuum manifold will be unable to evolve to
match up with one another without a significant energy input.  Such
field configurations are known as \emph{topological defects}, and the
idea that such configurations might arise in the natural evolution of
the Universe was first put forward by Kibble \cite{Kibble}.  The
existence of such solutions relies crucially on the topology of the
vacuum manifold; an arbitrary field theory will not, in general, allow
for topological defect solutions.

In the present work, I will address the question of whether
topological defect solutions can arise in theories with a tensor field
that spontaneously breaks Lorentz symmetry.  After some preliminaries
(Section \ref{prelimsec}), I discuss the topology of vacuum manifolds
of tensor fields in Section \ref{vacmansec}.  In Section
\ref{existsec}, I find maximally symmetric topological defect
solutions for those tensor fields that can support them; their basic
physical properties are described in Section \ref{physpropsec}.
Finally, I discuss more general issues arising from this work in
Section \ref{discsec}.

Throughout this work, we will use units in which $\hbar = c = 1$;
sign conventions for the metric and curvature will be those of Wald
\cite{Wald}.  In particular, the metric signature will be $(-, +, +,
+)$.  

\section{Preliminaries}
\label{prelimsec}

Most theories of current interest in which Lorentz symmetry is
spontaneously violated follow from an action of the form
\begin{equation}
  S = \int \dif^4 x \left( \frac{1}{2} \mt \cdot \mathcal{O}[\mt]
    - V(\mt) \right),
  \label{genaction}
\end{equation}
where $\mt$ is a tensor field of some rank $r$, potentially with some
symmetry relations; $\mathcal{O}$ is a linear, second-order,
self-adjoint, strictly differential operator on tensors of rank
$r$;\footnote{By self-adjoint, we mean here that $\mt_1 \cdot
  \mathcal{O}[\mt_2] = \mathcal{O}[\mt_1] \cdot \mt_2$ for all tensor
  fields $\mt_1$ and $\mt_2$, up to total derivatives.  By
  ``strictly'' differential, we mean that $\mathcal{O}$ only depends
  on the derivatives of $\mt$, and not on $\mt$ itself; any such
  dependence in $\mathcal{O}$ can simply be treated as part of the
  potential term.} and $V(\mt)$ is the potential energy for $\mt$.
(We assume for the moment a fixed flat background.)  The equation of
motion derived from an action of this form is then
\begin{equation}
  \mathcal{O}[\mt] - \frac{\delta V(\mt)}{\delta
    \mt} = 0.
  \label{genEOM}
\end{equation}

The potential term $V$ is a Lorentz scalar. Assuming that we do not
have any background geometric structure in this theory, this means
that $V$ must be a function of the various scalars that can be formed
out of $\mt$ via contraction of its indices with the metric
$\eta_{ab}$.  For example, if $\mt$ is an arbitrary two-index
tensor $t_{ab}$, we could have
\begin{equation}
  V(t_{ab}) = V(t_a {}^a, t_{ab} t^{ab}, t_{ab} t^{ba}, t_a {}^b t_b
  {}^c t_c {}^a, \dots).
\end{equation}
For the purposes of this paper, we will only consider potential terms
of the form
\begin{equation}
  V(\mt) = V(\mt^{a_1 a_2 \cdots a_r} \mt_{a_1 a_2 \dots a_r})
  \equiv V(\mt^{(a_i)} \mt_{(a_i)}),
\end{equation}
where we have introduced the notation $(a_i)$ to represent the index
string $a_1 a_2 \cdots a_r$.  (Where there is no risk of confusion, we
will use $(a)$ the same way.)  For a potential of this form, the
potential term in the equation of motion will just be
\begin{equation}
\frac{\delta V(\mt^{(a)})}{\delta \mt^{(a)}} = 2 V'(\mt^{(a)}
  \mt_{(a)}) \mt_{(a)}.
\end{equation}
This equation implies that if $\mt^{(a)}$ takes on a constant value
$\bar{\mt}^{(a)}$ everywhere in spacetime, such that
$V'(\bar{\mt}^{(a)} \bar{\mt}_{(a)}) = 0$, then the equation of motion
\eqref{genEOM} will be satisfied.  If the potential $V$ is constructed
in such a way that it is minimized at a non-zero value of its
argument, then $\mathcal{T}^{(a)}$ will be non-zero in this solution.

In this way, this model will spontaneously break Lorentz symmetry.
The above mentioned solutions of the equations of motion will not be
Lorentz-invariant: they contain a non-zero tensor field $\mt_{(a)} =
\bar{\mt}_{(a)}$ throughout spacetime, which imparts a preferred
geometric structure to flat spacetime.  Couplings between our
Lorentz-violating field and ``conventional'' matter fields (akin to
the hypothesized Yukawa coupling between fermion fields and the Higgs
field) could then give rise to a wide variety of observable physical
phenomena \cite{LVdata}.

It is important to note, however, that the specific value of
$\bar{\mt}^{(a)}$ is not uniquely determined by the equations of
motion; in fact, the action \eqref{genaction} is completely
Lorentz-invariant.  Rather, any constant tensor field satisfying
$\mt^{(a)} \mt_{(a)} = C$, where $V'(C) = 0$, will be a solution of
the equations of motion.  The set of all such tensors will form a
submanifold $\mvac$ in the space $\mathbb{V}$ of tensors under
consideration.  The shape of this manifold will be critical in
determining whether a tensor field taking values in $\mathbb{V}$ can
give rise to topological defects; it is to this question that we now
turn.

\section{Vacuum Manifolds}
\label{vacmansec}

The idea of topological defects in field theories is not a new one; a
thorough description of the idea in the context of high-energy physics
can be found in \cite{VilSh}, to which the interested reader is
referred.  For the purposes of this work, a topological defect can be
thought of as a solution of the equations of motion for which the
fields asymptotically approach a minimum-energy value as we go to
spatial infinity, but for which this minimum-energy value is dependent
on the direction that we go to infinity.  The types of topological
defects that can arise as solutions of a given theory will depend
critically on the topology of that theory's vacuum manifold $\mvac$.
If the vacuum manifold is disconnected, we can have \emph{domain wall}
solutions; if the vacuum manifold contains non-contractible loops,
\emph{cosmic strings} may arise; and if the vacuum manifold contains
non-contractible two-spheres, we can potentially have \emph{monopole}
solutions.  In terms of the homotopy groups of the manifold, such
structures will arise if the groups $\pi_0(\mvac)$, $\pi_1(\mvac)$, or
$\pi_2(\mvac)$ (respectively) are non-trivial.\footnote{This
  description excludes \emph{textures}, a type of non-localized
  topological defect solution that can arise when $\pi_3(\mvac)$ is
  non-trivial.  We will not explore these types of solutions in this
  work; see \cite{VilSh} for further details.}

To identify what types of topological defects can arise in our theory,
then, we need to know the topology of our vacuum manifold.  As noted
above, we will be concerned in this paper with the set of all
tensors $\mt^{(a)} \in \mathbb{V}$ with a fixed ``tensor norm'' given by
\begin{equation}
  \eta_{(a)(b)} \mt^{(a)} \mt^{(b)} = C,
  \label{tensornorm}
\end{equation}
where we have defined 
\begin{equation}
  \eta_{(a)(b)} = \eta_{a_1 b_1} \eta_{a_2 b_2} \cdots \eta_{a_r b_r}.
  \label{metricdef}
\end{equation}
$\mathbb{V}$ here is the space of all tensor of a definite rank and
symmetry type; since such sets are closed under addition and scalar
multiplication, we can view $\mathbb{V}$ as a real vector space.  The
tensor norm \eqref{tensornorm} defines a quadratic form (not
necessarily definite) on $\mathbb{V}$.  In the Appendix it is shown
that this quadratic form is non-degenerate for tensors of definite
rank and symmetry type non-degenerate.  This implies that we can pick
an orthonormal basis for $\mathbb{V}$, i.e., a set of $n$ tensors
$\{(e_i)^{(a)}\} \in \mathbb{V}$ such that
\begin{equation}
  \eta_{(a) (b)} (e_i)^{(a)} (e_j)^{(b)} = \pm \delta_{ij},
  \label{orthbasis}
\end{equation}
where $i$ and $j$ take on the values $\{ 1, \dots, n \}$, the plus
sign holds for the first $n_+$ of the basis elements, and the minus
sign holds for the remaining $n_-$.  In other words, $\mathbb{V}$ is
an $n$-dimensional real vector space on which we have a metric
$\eta_{(a)(b)}$ of signature $(n_+, n_-)$.  As the tensors
$\{(e_i)^{(a)}\}$ form a basis for $\mathbb{V}$, we can decompose an
arbitrary tensor $\mt^{(a)}$ in terms of components with respect to
this basis:
\begin{equation}
  \mt^{(a)} = \sum_{i=1}^n \mt_i \, (e_i)^{(a)}.
\end{equation}
We can then use the orthogonality properties \eqref{orthbasis} of our
basis to write the tensor norm \eqref{tensornorm} in terms of these
components:
\begin{equation}
  \eta_{(a)(b)} \mt^{(a)} \mt^{(b)} = \mt_1^2 + \mt_2^2 + \dots + \mt_{n_+}^2 -
  \mt_{n_+ +1}^2 - \dots - \mt_n^2.
  \label{componentnorm}
\end{equation}

Our vacuum manifold $\mvac$ will thus be the set of all tensors for
which this norm is a given constant $C$.  It is not hard to see that
$\mathbb{V}$ is equivalent to $\mathbb{R}^n$ (since the $\mt_i$'s can
be viewed as coordinates on $\mathbb{V}$), and that $\mvac$ will be
some $(n-1)$-dimensional hyperboloid embedded in $\mathbb{R}_n$.
Specifically, we note that if $C > 0$, our vacuum manifold will be the
space of all tensors whose components satisfy
\begin{equation}
   \mt_1^2 + \dots + \mt_{n_+}^2 = \mt_{n_+ +1}^2 + \dots + \mt_n^2 +
   C.
   \label{posnorm}
\end{equation}
This hyperboloid can be seen to be topologically equivalent to $S^{n_+
  -1} \times \mathbb{R}^{n_-}$: for any value of the $n_-$ components
$\mt_{n_+ + 1}$ through $\mt_n$, the components $\mt_1$ through
$\mt_{n_+}$ are constrained to lie on an $(n_+ - 1)$-sphere whose
radius squared is the right-hand side of \eqref{posnorm}.  Similarly,
for $C <0$, we can rearrange \eqref{componentnorm} to yield
\begin{equation}
   \mt_{n_+ +1}^2 + \dots + \mt_n^2 = \mt_1^2 + \dots + \mt_{n_+}^2 +
   (-C),
\end{equation}
which, by similar logic, yields a space that is topologically
equivalent to $S^{n_- -1} \times \mathbb{R}^{n_+}$.  Thus, in both
cases the vacuum manifold is homeomorphic to $S^p \times \mathbb{R}^q$
for some $p$ and $q$.\footnote{In the case where $C=0$, the set of
  ``null tensors'' in $\mathbb{V}$ can be shown to be the cone on
  $S^{n_+ - 1} \times S^{n_- - 1}$.  Since cone spaces are
  contractible, all of their homotopy groups are trivial, and so
  topological defects cannot arise.}

We can now see that the topology of $\mvac$, and thus the possibility
of topological defects in theories with spontaneous Lorentz breaking,
depends heavily on the signature of the metric $\eta_{(a)(b)}$ induced
on $\mathbb{V}$.  Since $\mathbb{R}^q$ is a contractible space for all
$q$, it follows that $\pi_N(S^p \times \mathbb{R}^q) = \pi_N(S^p)$ for
all $N$.  Moreover, since we are interested in topological defects, we
only need to consider the lower-dimensional homotopy groups $\pi_0$,
$\pi_1$, and $\pi_2$; for a $p$-sphere, these groups are non-trivial
if and only if $p = 0$, 1, or 2 respectively. Thus, the
lower-dimensional homotopy groups of our vacuum manifold $\mvac$ will
be trivial unless either $n_+$ or $n_-$ is less than or equal to 3; if
both are greater than 3, localized topological defects cannot arise.

\newlength{\youngpad}
\setlength{\youngpad}{3ex}
\begin{table}
  \begin{tabular}{lc|c|c}
    \multicolumn{2}{c|}{Rank \& } & \multicolumn{2}{c}{$(n_+,n_-)$} \\
    \multicolumn{2}{c|}{symmetry type} & \multicolumn{1}{c}{General} &
    Trace-free \\ 
    \hline
    $r = 0$ & $\emptyset$ & $(1,0)$ & $(1,0)$ \\[\youngpad] \hline
    $r = 1$ & $\yng(1)$ & $(3,1)$ & $(3,1)$ \\[\youngpad] \hline
    $r = 2$ & $\yng(2)$ & $(7,3)$ & $(6,3)$ \\[\youngpad]
            & $\yng(1,1)$ & $(3,3)$ & $(3,3)$ \\[\youngpad] \hline
    $r = 3$ & $\yng(3)$ & $(13,7)$ & $(10,6)$ \\[\youngpad]
            & $\yng(2,1)$ & $(11,9)$ & $(8,8)$ \\[\youngpad]
            & $\yng(1,1,1)$ & $(1,3)$ & $(1,3)$ \\[1.2\youngpad] \hline
    $r = 4$ & $\yng(4)$ & $(22,13)$ & $(15,10)$ \\[\youngpad]
            & $\yng(3,1)$ & $(24,21)$ & $(15,15)$ \\[\youngpad]
            & $\yng(2,2)$ & $(12,8)$ & $(5,5)$ \\[\youngpad]
            & $\yng(2,1,1)$ & $(6,9)$ & $(3,6)$ \\[1.2\youngpad]
            & $\yng(1,1,1,1)$ & $(0,1)$ & $(0,1)$ \\[1.4\youngpad] \hline
            & \rule{7em}{0pt} & \rule{5em}{0pt} & \rule{5em}{0pt}\\[-3ex]
    $r = 5$ & $\yng(5)$ & $(34,22)$ & $(21,15)$ \\[\youngpad]
            & $\yng(4,1)$ & $(45,39)$ & $(24,24)$ \\[\youngpad]
            & $\yng(3,2)$ & $(33,27)$ & $(12,12)$ \\[\youngpad]
            & $\yng(3,1,1)$ & $(15,21)$ & $(6,10)$ \\[1.2\youngpad]
            & $\yng(2,2,1)$ & $(11,9)$ & $(0,0)$ \\[1.2\youngpad]
            & $\yng(2,1,1,1)$ & $(1,3)$ & $(0,0)$
  \end{tabular}
  \caption{Signatures of the spaces of tensors of definite symmetry
    type with rank $r \leq 5$, with and without traces subtracted, in
    $d = 4$.} 
  \label{sigtable}
\end{table}
We have thus reduced the problem of determining the topology of the
vacuum manifold of a Lorentz-breaking tensor field to that of
determining the signature of the space $\mathbb{V}$ in which it lies.
These signatures can be determined by the iterative procedures
described in the Appendix; the results, for both general and
trace-free tensors of rank $r \leq 5$ and definite symmetry type
(labelled by Young tableaux), are given in Table \ref{sigtable}.  We
note that only three types of tensors (not counting the scalar) have
the correct vacuum topology to support topological defects in three
spatial dimensions:
\begin{itemize}
\item Vectors ($r=1$).  The space of vectors $v^a$ for which $v^a v_a
  = C$, with $C < 0$, has topology $S^0 \times \mathbb{R}^3$.  Note
  that $S^0$ is a set containing two discrete points; the vacuum
  manifold is thus the topologically equivalent to two disconnected
  copies of $\mathbb{R}^3$, which can be seen to be the past-oriented
  and future-oriented timelike vectors of a given norm.  Alternately,
  the space of vectors for which $v^a v_a = C > 0$ has topology $S^2
  \times \mathbb{R}$.  In principle, then, vectors that spontaneously
  break Lorentz symmetry could give rise to either domain wall or
  monopole solutions, depending on whether the vacuum manifold
  consists of timelike or spacelike vectors respectively.
\item Antisymmetric two-tensors.  The space of such tensors $B^{ab}$
  for which $B^{ab} B_{ab} = C$ will have topology $S^2 \times
  \mathbb{R}^3$ for any non-vanishing $C$ (positive or negative.)
  Thus, these tensors could in principle give rise to monopoles.
\item Symmetric two-tensors, with or without trace.  The space of such
  tensors $h^{ab}$ for which $h^{ab} h_{ab} = C < 0$ will have
  topology $S^2 \times \mathbb{R}^7$ (or $S^2 \times \mathbb{R}^6$ for
  trace-free tensors.)  Such tensors could then in principle give rise
  to monopoles as well.
\end{itemize} 
The remaining tensors in Table \ref{sigtable} with low signatures
(i.e. $n_+ \leq 3$ or $n_- \leq 3$) can be shown to be equivalent to
either the scalar or one of the three types of tensors above; see the
Appendix and \cite{Hamermesh} for details.

\section{Existence of topological defect solutions}
\label{existsec}

In the previous section, we found that the vacuum manifolds $\mt^{(a)}
\mt_{(a)} = C$ of only three types of tensors (vectors, antisymmetric
two-tensors, and symmetric two-tensors) have the appropriate topology
to support topological defects.  All other tensors with rank $r \leq
5$ either have both $n_+$ and $n_-$ too large to support topological
defects in three spatial dimensions, or are equivalent to one of these
three types of tensors.

While the existence of a vacuum manifold of the proper topology is a
necessary condition for the existence of topological defect solutions,
it is not a sufficient condition.  By a topological defect solution,
we mean a solution of the equation of motion 
\begin{equation}
  \mathcal{O}_{(a)}[\mt] - 2 V'(\mt^{(b)} \mt_{(b)}) \mt_{(a)} = 0
\end{equation}
such that the tensor field $\mt^{(a)}$ goes asymptotically to its
vacuum manifold, and such that the asymptotic map between ``spatial
infinity'' (in the appropriate sense) and the vacuum manifold is
topologically non-trivial.  While such asymptotic maps are easily
constructed, the global existence of such solutions will also depend
on the properties of the kinetic operator $\mathcal{O}$.  (See
\cite{Babi} for an example of this in the scalar case.)  We must thus
ask what type of operators are appropriate for our theories.

Fortunately, for the three types of tensor fields under consideration,
there are already ``natural'' choices of kinetic term.  For the vector
and anti-symmetric two-tensor cases, we can choose a
``field strength--squared'' kinetic term.  Specifically, for the vector
field $A^a$, we can write
\begin{equation}
  S_A = \int \dif^4 x \left( - \frac{1}{4} F^{ab} F_{ab} - V(A^a A_a)
  \right),
  \label{vecaction}
\end{equation}
where $F_{ab} = 2 \partial_{[a} A_{b]}$.  For the antisymmetric
two-tensor $B_{ab}$, we can write
\begin{equation}
  S_B = \int \dif^4 x \left( - \frac{1}{6} F^{abc} F_{abc} - V(B^{ab}
    B_{ab}) \right),
  \label{ASaction}
\end{equation}
where $F_{abc} = 3 \partial_{[a} B_{bc]}$.  (A recent study of this
and related models can be found in \cite{LVAS}.)  For the symmetric
two-tensor, meanwhile, we can take the standard kinetic operator
$\mathcal{K}$ for a massless spin-2 field:
\begin{equation}
  S_\gamma = \int \dif^4 x \left( \frac{1}{2} \gamma^{ab}
    \mathcal{K}_{abcd} \gamma^{cd} 
    - V(\gamma^{ab} \gamma_{ab}) \right)
  \label{symaction}
\end{equation}
where
\begin{multline}
  \mathcal{K}_{abcd} = \frac{1}{2} \left[ (\eta_{a(c} \eta_{d)b} -
    \eta_{ab} \eta_{cd}) \Box + \eta_{ab} \partial_c \partial_d
  \right. \\
  \left. + \eta_{cd} \partial_a \partial_b -
    \eta_{a(c} \partial_{d)} \partial_b -
    \eta_{b(c} \partial_{d)} \partial_a \right].
\end{multline}

The exact field profile of any topological defect solutions will also
depend on the form of the tensor potential $V(x)$.  However, the
existence of such solutions is in general largely independent of the
exact functional form of $V(x)$; as long as $V(x)$ has a single
minimum at a value of $x$ of the appropriate sign, any other changes
to $V(x)$ will just change the fine details of the field profile of
the defect solution.  For concreteness, we will take $V(x)$ to in all
three cases be of the form
\begin{equation}
  V(x^2) = \frac{\lambda}{2} (x^2 \pm b^2)^2,
\end{equation}
where the sign is chosen depending on whether we want the topological
defect to arise from the negative-norm components of our tensor or the
positive-norm components.  Note that in the case of a scalar field,
this reduces to the familiar fourth-order double-well potential.

We are now in a position to look for topological defect solutions for
tensor fields of the three types above.  We will treat these cases
below, in order of increasing physical interest.

\subsection{Symmetric tensors}

In the case of symmetric two-tensors, we have $(n_+, n_-) = (7,3)$ (or
$(6,3)$ if we require the trace to vanish.)  Since $n_+ > 3$ and
$n_-=3$, a topological defect solution (if it exists) will be a
monopole, and the vacuum manifold will be of the form $\gamma^{ab}
\gamma_{ab} = - b^2$.  The Euler-Lagrange equation derived from the
action \eqref{symaction} is
\begin{equation}
  \label{symEOM}
  \mathcal{E}_{ab} \equiv 
  \mathcal{K}_{ab} {}^{cd} \gamma_{cd} - 2 \lambda (\gamma_{cd} \gamma^{cd} +
  b^2) \gamma_{ab} = 0
\end{equation}
The simplest possible topological defect solution will be one
possessing spherical symmetry.  This constricts the form that
$\gamma_{ab}$ can take;  the spherical coordinate components
$\gamma_{\mu \nu}$ must be of the form
\begin{align}
  \gamma_{tt} &= f(r), & \gamma_{rt} &=
  \gamma_{tr} = h(r),  \\
 \gamma_{rr} &= g(r), & \gamma_{\theta \theta} &= \sin^{-2} \theta
 \gamma_{\phi \phi} = r^2  i(r).
\end{align}
In terms of these functions, the independent coordinate components
$\mathcal{E}_{\mu \nu}$ of the equations of motion are then
\begin{subequations}
\begin{equation}
  \mathcal{E}_{tt} = i'' + \frac{3}{r} i' + \frac{1}{r^2} i -
  \frac{1}{r} g' - \frac{1}{r^2} g + 2 \lambda (\gamma^2 + b^2) f = 0,
\end{equation}
\begin{equation}
  \mathcal{E}_{tr} = 2 \lambda (\gamma^2 + b^2) h = 0,
  \label{hvanish}
\end{equation}
\begin{equation}
  \mathcal{E}_{rr} = \frac{1}{r} f' + \frac{1}{r^2} g - \frac{1}{r} i'
  - \frac{1}{r^2} i + 2 \lambda (\gamma^2 + b^2) g = 0,
\end{equation}
and
\begin{multline}
  \frac{1}{r^2} \mathcal{E}_{\theta \theta} = \frac{1}{2} \left( i'' -
      \frac{2}{r} i' + f'' + \frac{1}{r} f' +
  \frac{1}{r} g'\right) \\ + 2 \lambda (\gamma^2 + b^2) i = 0,
\end{multline}
\end{subequations}
where $\gamma^2 \equiv f^2 + g^2 + 2 i^2 - 2 h^2$ and primes denote
derivatives with respect to $r$.  It is evident from equation
\eqref{hvanish} that we must either have $\gamma^2 + b^2 = 0$
everywhere in spacetime---in which case our field is everywhere
confined to the vacuum manifold---or we must have $h = 0$ everywhere,
in which case the field cannot asymptotically approach the vacuum
manifold.  Thus, we conclude that the symmetric two-tensor action
\eqref{symaction} cannot support spherically symmetric topological
defect solutions.\footnote{Requiring $\gamma_{ab}$ to be trace-free
  merely sets $i = \frac{1}{2}(f-g)$, and does not affect the above
  argument.}

\subsection{Vectors}

In the case of vector fields, we have $(n_+, n_-) = (3,1)$;  thus, we
could either have monopole solutions (if $C = b^2$) or domain wall
solutions (if $C = -b^2$.)  The Euler-Lagrange equation derived from
the action \eqref{vecaction} is
\begin{equation}
  \label{vecEOM}
  \mathcal{E}_a \equiv \partial^b F_{ba} - 2 \lambda (A_b A^b \pm b^2) A_a
  = 0.
\end{equation}
Since the expected symmetry of the simplest topological defect
solutions is different for spacelike and timelike vectors (spherical
and planar, respectively), we must treat these cases separately.

\subsubsection{Spacelike vacuum manifold}

If the vacuum manifold consists of all vectors $A^a$ with $A^a A_a =
b^2$, our topological defect solution (if it exists) will be a
monopole solution;  thus, we look for solutions with spherical symmetry.
The most general vector field with such a symmetry will be
\begin{equation}
(A^t, A^r, A^\theta, A^\phi) = (f(r), g(r), 0, 0).
\end{equation}
This implies that the field strength tensor $F_{ab}$ has components
$F_{tr} = - F_{rt} = f'$, with all other components vanishing; the
components of the equation of motion then become
\begin{subequations}
\begin{equation}
  \mathcal{E}_t = - f'' - \frac{2}{r} f' + 2 \lambda(- f^2 + g^2 - b^2)
  f = 0
\end{equation}
and
\begin{equation}
  \mathcal{E}_r = - 2 \lambda(- f^2 + g^2 - b^2) g = 0.
  \label{gvanish}
\end{equation}
\end{subequations}
(Primes again denote differentiation with respect to $r$.)  We see
from equation \eqref{gvanish} that as in the symmetric tensor case,
the field must either be in the vacuum manifold everywhere in
spacetime or $g$ must vanish everywhere;  neither case can correspond
to a topological defect.

\subsubsection{Timelike vacuum manifold}

The case where the vacuum manifold consists of all vectors $A^a$ with
$A^a A_a = - b^2$ is somewhat more promising.  In this case, we have
the possibility of domain wall solutions.  The simplest possible
domain wall will have planar symmetry, with all fields depending on
some Cartesian coordinate $x$;  the vector field will take on the form
\begin{equation}
  (A^t, A^x, A^y, A^z) = (f(x), g(x), 0, 0).
\end{equation}
Using primes here to denote differentiation with respect to $x$, we
have $F_{tx} = - F_{xt} = f'$;  the non-trivial components of the
equation of motion \eqref{vecEOM} are 
\begin{subequations}
\begin{equation}
  \mathcal{E}_t = - f'' + 2 \lambda( -f^2 + g^2 + b^2) f = 0
  \label{DWeq}
\end{equation}
and
\begin{equation}
  \mathcal{E}_x = - 2 \lambda( - f^2 + g^2 + b^2) g = 0
  \label{vecEOMx}
\end{equation}
\end{subequations}
As in the spacelike vector case, we see from \eqref{vecEOMx} that if
we do not want the solution to lie in the vacuum manifold everywhere,
the spacelike component of $A^a$ (namely $g$) must vanish.  However,
in this case the vanishing of $g$ is not an impediment to the field
asymptotically approaching the vacuum manifold.  In fact, the equation
\eqref{DWeq} can be seen to be exactly that of the prototypical domain
wall solution, arising in a theory of a single scalar field with a
broken $\mathbb{Z}_2$ symmetry (see, for example, Chapter 3 of
\cite{VilSh}).  Its solution is
\begin{equation}
  \label{DWsol}
  f(x) = \pm b \tanh \left( \sqrt{-\lambda} b x \right).
\end{equation}
In this case, we find that a ``Lorentz-violating'' topological defect
solution does exist; it is a domain wall configuration, with the
vector field past-oriented on one side of the wall, future-oriented on
the other, and smoothly interpolating through $A^a = 0$ in between.
Notably, if this solution is to exist, we must have $\lambda
< 0$; this will turn out to be quite important when we analyse the
properties of this solution in Section \ref{DWpropsec}.

\subsection{Antisymmetric tensors}

\begin{figure}
  \includegraphics[width=\figwidth]{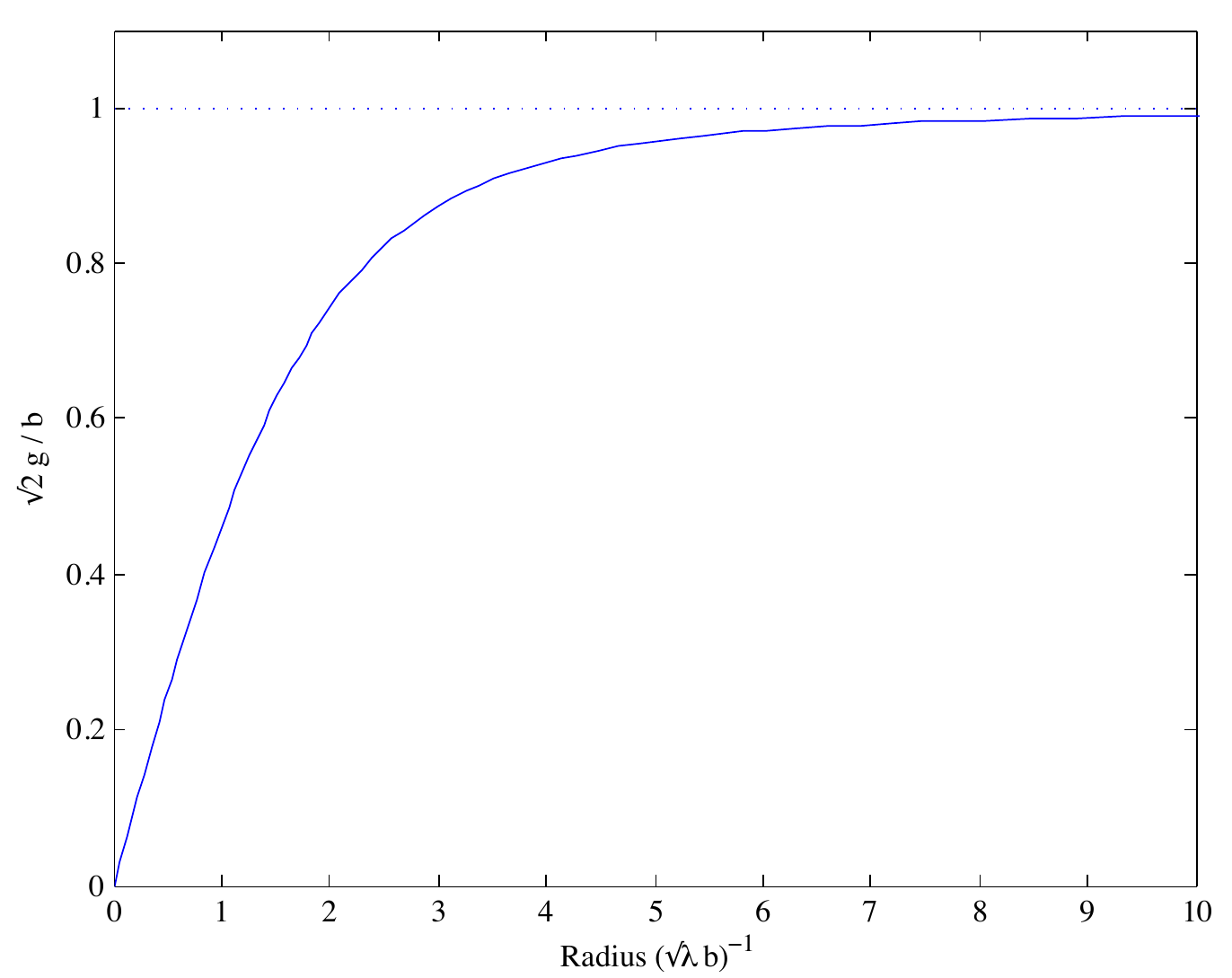}
  \caption{\label{ASfieldfig} Field configuration for antisymmetric
    tensor monopole solution.  For this solution, $B_{\theta \phi} = -
    B_{\phi \theta} = g(r) r^2 \sin \theta$; all other components of
    $B_{ab}$ vanish.}
\end{figure}

In the case of antisymmetric two-tensors, we have $(n_+, n_-) =
(3,3)$.  This implies that monopole solutions are topologically
allowed for both positive-norm and negative-norm tensors.  Since we
expect both types of tensors to have spherically symmetric solutions,
we can treat both cases simultaneously.  The Euler-Lagrange equation
derived from the action \eqref{ASaction} is
\begin{equation}
  \label{ASEOM}
  \mathcal{E}_{ab} \equiv \partial^c F_{cab} - 2 \lambda (B^{cd}
  B_{cd} \pm b^2) B_{ab} = 0.
\end{equation}
The most general antisymmetric two-tensor with spherical symmetry can
be written in the form
\begin{align}
  \label{Bcomps}
  B_{tr} = - B_{rt} &= f(r), & B_{\theta \phi} = - B_{\phi \theta} &=
  g(r) r^2 \sin \theta,
\end{align}
with all other components vanishing.  There are then two non-trivial
components of the equation of motion:
\begin{subequations}
\begin{equation}
  \mathcal{E}_{tr} = -2 \lambda (- 2f^2 + 2g^2 \pm b^2) f = 0,
\end{equation}
and
\begin{equation}
  \mathcal{E}_{\theta \phi} = \frac{\partial}{\partial r}\left(
    \frac{\partial g}{\partial r} +
    \frac{2}{r} g \right) - 2 \lambda (-2 f^2 + 2 g^2 \pm b^2 ) g = 0.
\end{equation}
\end{subequations}
Assuming again that we do not want a solution where the field is
everywhere in its vacuum manifold, we must have $f = 0$ from the first
equation above.  This then implies that the vacuum manifold must
consist of positive-norm (``spacelike'') tensors; we therefore choose
the minus sign in \eqref{ASEOM} so that $C = b^2$.  Defining rescaled
variables $\tg$ and $\tr$ such that $g = b \tg / \sqrt{2}$ and $r =
\tr/ (\sqrt{2 \lambda} b)$, the second equation becomes
\begin{equation}
  \label{tgeqn}
  \frac{\partial}{\partial \tr} \left(\frac{ \partial
      \tg}{\partial \tr} +
    \frac{2}{\tr} \tg \right) - (\tg^2 - 1) \tg = 0.
\end{equation}
This equation and its solutions were briefly discussed in
\cite{lormon}.  Up to rescaling, Equation \eqref{tgeqn} is exactly the
differential equation that arises in a ``hedgehog monopole'' solution
\cite{BarrVil}, in which the spontaneously broken symmetry is an
internal $O(3)$ symmetry among a triplet of Lorentz scalars.  While a
closed-form analytic solution for $g$ is not known, we can use
numerical integration or series techniques \cite{ShiLi, HarLou} to
obtain the form of $g$ (Figure \ref{ASfieldfig}.)
We can also expand $\tg$ as a power series in $1/\tr$ to
examine its asymptotic behaviour;  the result is
\begin{equation}
  \label{gasymp}
  \tg(\tr) = 1 - \frac{1}{2\tr^2} -\frac{3}{2 \tr^4} + \dots.
\end{equation}
We also note that the equation \eqref{tgeqn} is invariant under the
transformation $\tg \to - \tg$.  A solution that asymptotically
approaches $g = -b/\sqrt{2}$ rather than $g = b/\sqrt{2}$ can be
thought of as an antimonopole rather than a monopole solution.  

\section{Physical properties of Lorentz defect solutions}
\label{physpropsec}

In the previous section, we found that static, maximally symmetric
topological defect solutions exist for two types of tensor fields:
vectors and antisymmetric two-tensors.  A natural question to ask
concerning these solutions regards the form of their stress-energy;
the gravitational effects of such solutions would be a natural (and,
in the absence of an explicit coupling between these fields and
``conventional'' matter fields, the only) way to detect them.

\subsection{Vector domain walls}
\label{DWpropsec}

In the case of vector fields, we were able to write down an exact
solution \eqref{DWsol} representing a domain-wall solution of the
equations of motion \eqref{vecEOM}.  The stress-energy tensor
associated with the vector field $A^a$ can be found be the usual
technique of differentiating the action \eqref{vecaction} with respect
to the metric:
\begin{multline}
  \label{vecSET}
  T_{ab} = F_{ac} F_b {}^c - \frac{1}{2} \eta_{ab} F_{cd} F^{cd} \\ +
  \lambda \left( (A^c A_c + b^2) A_a A_b - \frac{1}{2} \eta_{ab} (A^c A_c +
  b^2)^2 \right).
\end{multline}
Note the presence of the third term here, which arises due to the
differentiation of our potential $V(A_a A_b g^{ab})$ with respect
to the metric.  Terms such as this do not arise for topological
defects constructed out of Lorentz scalars;  the fact that our fields
are Lorentz tensors requires that our potential depend on the metric
as well. 

Plugging our solution \eqref{DWsol} into \eqref{vecSET} yields
\begin{subequations}
  \label{DWSET}
\begin{align}
  T_{tt} &= \lambda b^4 \tanh^2( \sqrt{-\lambda} b x) \sech^2(
  \sqrt{-\lambda} b x), \\
  T_{xx} &= 0, \\
  T_{yy} = T_{zz} &= \lambda b^4 \sech^4 ( \sqrt{-\lambda} b x).  
\end{align}
\end{subequations}
We can take a ``thin-wall'' limit of this solution by requiring that
the wall thickness $(\sqrt{-\lambda} b)^{-1}$ go to zero while the
surface energy density and tension of the wall be held constant.
These latter two quantities are given by
\begin{equation}
\sigma = \int \dif x \, T_{tt} = - \frac{2}{3} \sqrt{-\lambda} b^3 
\end{equation}
and
\begin{equation}
\tau = - \int \dif x \, T_{yy} = \frac{4}{3} \sqrt{-\lambda} b^3 
\end{equation}
respectively.

We can now see an important aspect of our domain wall solution: its
surface energy density is negative.  This stems from the requirement
for existence of this solution, noted above, that we take the constant
$\lambda$ to be negative rather than positive.  The gravitational
dynamics of thin domain walls with a general surface density and
tension were examined by Ipser and Sikivie \cite{IpsSik}.  In
particular, it is shown that a domain wall is ``attractive'' if
$\sigma - 2 \tau > 0$, and ``repulsive'' if $\sigma - 2 \tau < 0$;
more precisely, two observers on opposite sides of the wall must
accelerate outwards to remain a constant distance apart if $\sigma - 2
\tau > 0$, and must accelerate inwards to remain a constant distance
apart if $\sigma - 2 \tau < 0$.  Since our vector domain walls have
$\sigma - 2 \tau < 0$, they fall into the latter, ``repulsive''
category.

However, there are some troubling aspects of this solution.  Since we
were forced to set $\lambda < 0$ to allow existence of the solution
\eqref{DWsol}, we can see from \eqref{DWSET} that the energy density
$\rho = T_{tt}$ and transverse pressure $P_\perp = T_{yy} = T_{zz}$
are negative.  This implies, in particular, that this solution does
not satisfy any of the standard energy conditions (weak, null, strong,
or dominant.)

A more serious problem concerns the stability of this theory.
Choosing $\lambda < 0$ means that the potential $V(A^a A_a)$ is
unbounded below rather than unbounded above.  In essence, we have
inverted the potential; instead of the vacuum manifold lying at the
bottom of the brim of a ``Mexican hat'', it is instead perched atop a
``Bundt cake.''  It thus seems likely that a small perturbation will
cause our fields to ``roll down the hill,'' and thus that this domain
wall will be unstable.

Further evidence for this can be found by obtaining the Hamiltonian
associated with the action \eqref{vecaction} for a general potential
$V(A^a A_a)$.  Taking the field variables to be the components of
$A_a$, and denoting the spatial components of $A_a$ with Roman indices
$i,j,k,\dots$, the Hamiltonian can be shown to be
\begin{multline}
  H = \int \dif^3 \vec{x} \bigg[ \frac{1}{2} \Pi^i \Pi_i + \frac{1}{4}
    F^{ij} F_{ij} + V(-(A_0)^2 + \vec{A}^2) \\ + 2 V'(-(A_0)^2 +
    \vec{A}^2) (A_0)^2 + \Pi^0 u \bigg],
\end{multline}
where the conjugate momenta $\Pi^i \equiv F^{i0}$, $u$ is a Lagrange
multiplier, and the fields are subject to the constraints
\begin{equation}
  \Pi^0 = 0
\end{equation}
and
\begin{equation}
  \partial_i \Pi^i + 2 V'(A^2) A_0 = 0.
\end{equation}
If the function $V(x)$ is unbounded below for positive values of its
argument $x$, this Hamiltonian can be seen to be unbounded both above
and below: it can be made arbitrarily positive by taking $A_0 \to 0$,
$A_i \to 0$, and $\Pi^i$ to be large but divergence-free, and can be
made arbitrarily negative by taking $A_0 \to 0$, $\Pi^i \to 0$, and
$A_i$ large and slowly varying.  This implies that the magnitudes of
our fields are not bounded by energy conservation; we must view this
as strong evidence that the full field theory \eqref{vecaction} with
$\lambda < 0$ is unstable, and (if so) physically unrealistic.

\subsection{AS tensor monopoles}

In the case of antisymmetric two-tensor fields, we found a numerical
solution (shown in Figure \ref{ASfieldfig}) which represents a
monopole solution of the theory with the action \eqref{ASaction}.  The
physical properties of this solution were briefly discussed in
\cite{lormon}; we review and elaborate upon this work here.

The stress-energy tensor associated with the action \eqref{ASaction}
is
\begin{multline}
  \label{ASSET}
  T_{ab} = F_{acd} F_b {}^{cd} - \frac{1}{6} \eta_{ab}
  F_{cde} F^{cde} \\ + \lambda \left( 4 (B_{de} B^{de} - b^2) B_{ac} B_b
    {}^c - \frac{1}{2} \eta_{ab} (B_{cd} B^{cd} - b^2)^2 \right).
\end{multline}
In terms of the function $g(r)$ defined in \eqref{Bcomps}, the
the energy density $\rho = T_{tt}$,
the radial pressure $P_r = T_{rr}$, and the tangential pressure
$P_\theta = r^{-2} T_{\theta \theta}$ are
\begin{subequations}
\begin{align}
  \rho &= \left( g' + \frac{2}{r} g \right)^2 +
  \frac{\lambda}{2} (2 g^2 - b^2)^2, \\
  P_r &= \left( g' + \frac{2}{r} g \right)^2 - \frac{\lambda}{2} (2
  g^2 - b^2)^2, \\
  P_\theta &=
  \left( g' + \frac{2}{r} g \right)^2 + 
  \frac{\lambda}{2}\left( 8g^2 (2g^2 - b^2) - (2g^2 - b^2)^2 \right).
\end{align}
\end{subequations}

\begin{figure}
  \includegraphics[width=\figwidth]{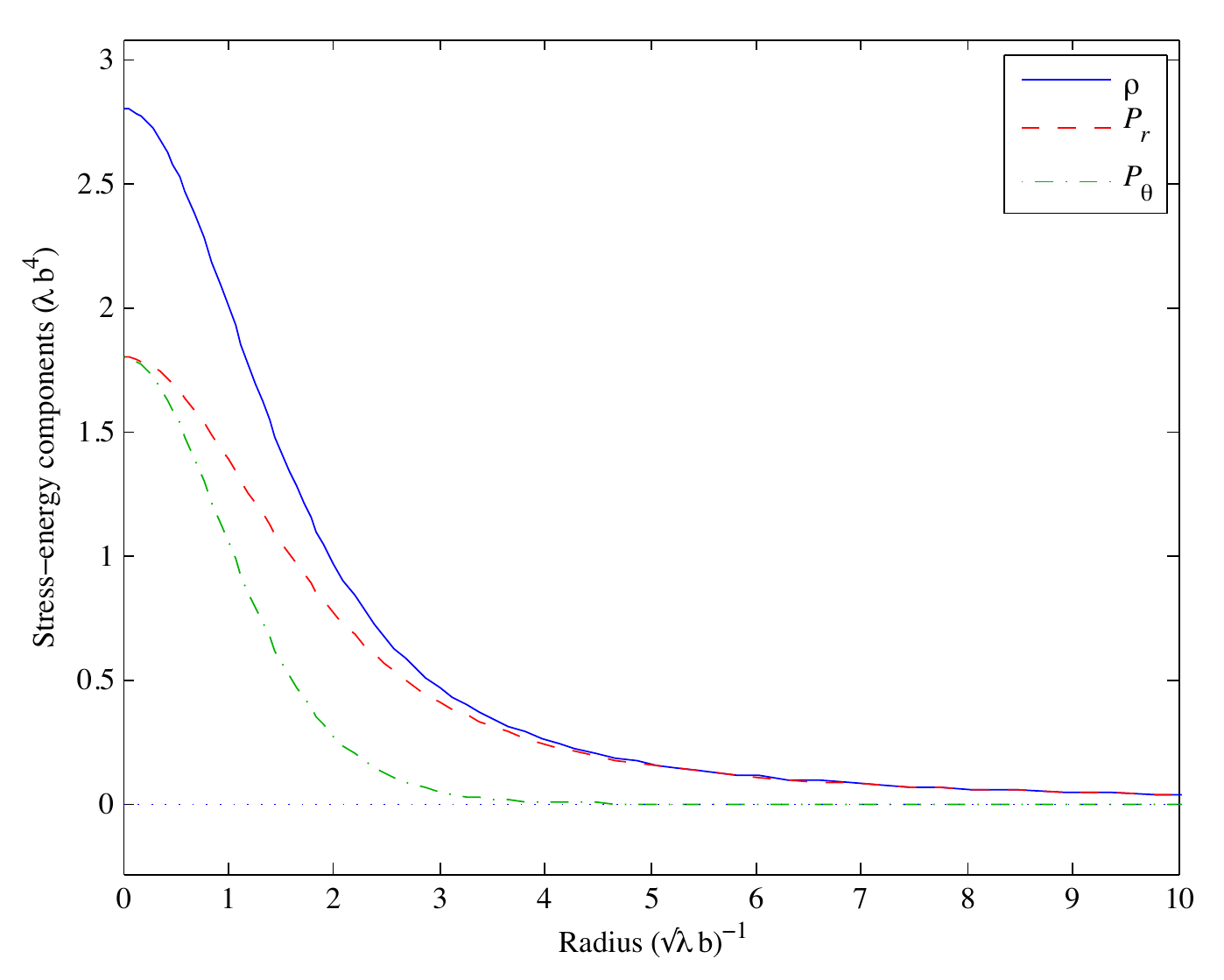}
  \caption{\label{ASSETfig} Energy density $\rho$, radial pressure
    $P_r$, and tangential pressure $P_\theta$ for the antisymmetric
    tensor monopole.  Note the positivity of all three quantities.}
\end{figure}

It is important to note that while the function $g(r)$ satisfies the
same equation of motion as the $O(3)$ scalar monopole described in
\cite{BarrVil}, the form of the stress-energy tensor is rather
different.  In particular, the $O(3)$ scalar monopole has positive
energy density but negative pressures (both radial and tangential),
while for our monopole all three quantities ($\rho$, $P_r$, and
$P_\theta$) are everywhere positive (see Figure \ref{ASSETfig}.)
We can use the asymptotic form of $\tilde{g}$ from Equation
\eqref{gasymp} to see the fall-off properties of the stress-energy
components; these work out to be
\begin{subequations}
\begin{equation}
  \rho = \lambda b^4 \left[ \frac{4}{\tr^2} + \frac{1}{\tr^4} + \dots
  \right],
\end{equation}
\begin{equation}
  P_r = \lambda b^4 \left[ \frac{4}{\tr^2} - \frac{1}{\tr^4} + \dots
  \right],
\end{equation}
and
\begin{equation}
  P_\theta = \lambda b^4 \left[ \frac{1}{\tr^4} + \dots \right].
\end{equation}
\end{subequations}
Note that the fall-off rate of $P_\theta$ is significantly faster than
that of $\rho$ and $P_r$;  this is due to an exact cancellation
between the $\mathcal{O}(r^{-2})$ dependencies of the kinetic portion
and the potential portion of the stress-energy tensor.

This difference in sign between the pressures of the $O(3)$ monopole
and our tensor monopole will lead to significant differences when we
examine the gravitational effects of this field configuration.  In the
case of a dynamical metric $g_{ab}$, the tensor equations of motion
are
\begin{subequations}
\begin{multline}
  (\mathcal{E}_G)_{ab} \equiv G_{ab} - 8 \pi G \left[ F_{acd} F_b {}^{cd} -
    \frac{1}{6} g_{ab} F_{cde} F^{cde} \right. \\ 
  + \lambda \bigg( 4 (B_{de}
      B^{de} - b^2) B_{ac} B_b 
      {}^c \\ \left. 
      - \frac{1}{2} \eta_{ab} (B_{cd} B^{cd} - b^2)^2 \bigg)
  \right] = 0
\end{multline}
and
\begin{equation}
  (\mathcal{E}_B)_{ab} \equiv \nabla^c F_{cab} - 2 \lambda (B_{cd} B^{cd} -
  b^2) B_{ab} = 0.
\end{equation}
\end{subequations}
Since we are assuming spherical symmetry and staticity, we can use
Schwarzschild coordinates to write our line element as
\begin{equation}
  \label{metricform}
  ds^2 = - M^2(r) dt^2 + N^2(r) dr^2 + r^2 (d \theta^2 + \sin^2
  \theta d \phi^2 ).
\end{equation}
We can use the same ansatz \eqref{Bcomps} as we did in flat spacetime
for the $\theta \phi$-component of $B_{ab}$, and take $B_{tr}$ to
vanish.  In terms of our ansatz functions $M(r)$, $N(r)$, and $g(r)$,
these equations become
\begin{subequations}
  \label{MNgEOMs}
\begin{multline}
  \frac{2}{\tr} \frac{N'}{N} + \frac{1}{\tr^2} (N^2 - 1) \\= 4 \pi G
  b^2 \left( \left( \tg' + \frac{2}{\tr} \tg \right)^2 +
    \frac{1}{2} N^2 (\tg^2 - 1)^2 \right), \label{EttEOM}
\end{multline}
\begin{multline}
  \frac{2}{\tr} \frac{M'}{M} - \frac{1}{\tr^2} (N^2 - 1) \\= 4 \pi G
  b^2 \left( \left( \tg' + \frac{2}{\tr} \tg \right)^2 - \frac{1}{2}
    N^2 (\tg^2 - 1)^2 \right), \label{ErrEOM}
\end{multline}
and
\begin{multline}
  \frac{\partial}{\partial \tr} \left( \tg' + \frac{2}{\tr} \tg
  \right) + \left( \frac{M'}{M} - 
    \frac{N'}{N} \right) \left( \tg' + \frac{2}{\tr} \tg \right) \\ - N^2
  (\tg^2 - 1) \tg = 0, \label{EBEOM}
\end{multline}
\end{subequations}
where we have rescaled our coordinates and fields as in the flat
spacetime case; primes denote differentiation with respect to
$\tr$.\footnote{These three equations \eqref{MNgEOMs} are proportional
  to $(\mathcal{E}_G)_{tt}$, $(\mathcal{E}_G)_{rr}$, and
  $(\mathcal{E}_B)_{\theta \phi}$, respectively; the equation arising
  from $(\mathcal{E}_G)_{\theta \theta}$ is non-trivial, but is
  automatically satisfied via the Bianchi identities as long as the
  other three equations hold.}

These equations do not have an obvious closed-form solution.  However,
we can obtain some interesting information concerning the asymptotic
properties of these solutions by taking the BPS limit
\cite{Bogo,PrasSomm}, in which we set $\tg \to 1$ exactly; this
corresponds to taking $\lambda \to 0$ while still looking for
solutions with the asymptotic behaviour of a topological defect.  In
this limit, the components of the Einstein equation \eqref{EttEOM} and
\eqref{ErrEOM} become
\begin{align}
  \frac{2}{\tr} \frac{N'}{N} + \frac{1}{\tr^2} (N^2 - 1) &\approx
  \frac{\epsilon}{\tr^2}, \\
  \frac{2}{\tr} \frac{M'}{M} - \frac{1}{\tr^2} (N^2 - 1) &\approx
  \frac{\epsilon}{\tr^2}, 
\end{align}
where we have defined $\epsilon \equiv 16 \pi G b^2$.  These equations
have the exact solution
\begin{align}
  M^2(r) &= C_2 \frac{\tr^{1 + \epsilon} + C_1}{\tr^{1-\epsilon}}, \\
  N^2(r) &= (1 + \epsilon) \frac{ \tr^{1 + \epsilon}}{\tr^{1 + 
      \epsilon} + C_1},
\end{align}
where $C_1$ and $C_2$ are constants of integration.  (Note that $C_2$
can be set to an arbitrary positive value via a rescaling of $t$;  we
will henceforth set $C_2 = 1$.) 

It is important to note that the stress-energy fall-off properties of
this solution are not quite the same as those in flat spacetime.
Asymptotically, the energy density $\rho$ and radial pressure $P_r$
are given by
\begin{equation}
  \rho \approx P_r \approx N^{-2} \frac{4 g^2}{r^2} \approx
  \frac{4 \lambda b^4}{\tr^2},
\end{equation}
which are the same as in the flat spacetime case.  However, the
fall-off rate of the tangential pressure $P_\theta$ must change; in
spherical coordinates, the Bianchi identity implies
that\footnote{Simply plugging in the BPS approximation $g =
  b/\sqrt{2}$ into the appropriate expression for $P_\theta$ yields
  $P_\theta \approx \frac{4 \lambda b^4}{\tr^2}$, which is
  inconsistent with the Bianchi identity.  From our experience in the
  flat case, we recall that the potential term also contributes in a
  critical way to the asymptotic behaviour of $P_\theta$; this
  contribution is fundamentally inaccessible to the BPS approximation,
  and illustrates its limitations in this situation. In a more careful
  analysis (expanding $M'/M$, $N$, and $g$ in a power series in
  $r^{-1}$), we see that in curved spacetime the kinetic and potential
  pieces do not quite cancel at $\mathcal{O}(r^{-2})$ when the metric
  is curved, rather than exactly cancelling as they did in a fixed
  flat background. }
\begin{equation}
  \nabla^\mu T_{\mu r} = \frac{\partial P_r}{\partial r} +
  \frac{M'}{M}(\rho + P_r) + \frac{2}{r} (P_r - P_\theta) = 0
\end{equation}
Plugging in our asymptotic expressions for $\rho$, $P_r$, and $M$, we
see that in curved spacetime $P_\theta$ must have the asymptotic form
\begin{equation}
  P_\theta \approx \frac{\epsilon}{2}(\rho + P_r) \approx \frac{4
    \epsilon \lambda b^4}{\tr^2}.
\end{equation}
If our mass scale $b$ is significantly smaller than the Planck mass,
we will still have $P_\theta \ll \rho$ and $P_\theta \ll P_r$
asymptotically, as we did in flat spacetime.  However, the
introduction of a curved metric requires that the fall-off
rate of $P_\theta$ change from $r^{-4}$ to $r^{-2}$.

Returning to the case of non-zero $\lambda$, we expect that our
solution will still have the same asymptotic behaviour, with $M(r)
\propto \tr^\epsilon$ and $N(r) \to \sqrt{1 + \epsilon}$.  The fact
that $N$ does not go to unity as $r \to \infty$ implies that the
constant-time slices of this spacetime have a spherical deficit angle;
specifically, the equatorial plane (i.e., $\theta = \pi/2$) has the
asymptotic geometry of a cone with a deficit angle $\pi \epsilon$.
The divergence of the $g_{tt}$ component of the metric might seem to
be a cause for concern; it does not appear that, for example, such a
metric could be asymptotically flat.  For a monopole in isolation,
this could imply that the full solution is inherently non-static; an
analogous situation would be that of anti-de Sitter space in
Schwarzschild coordinates, in which the $tt$-component of the metric
diverges proportionally to $r^2$.  It can also be shown that various
curvature invariants of our metric go to zero as $r\to \infty$; the
Ricci scalar is proportional to $r^{-2}$ (as might be expected from
the stress-energy tensor), and the squares of the Ricci tensor and the
Riemann tensor both fall off as $r^{-4}$.

Moreover, in a realistic physical situation, we only expect this
solution to be valid out to some finite radius where the effects of
larger structure (on galactic or cosmic scales) take over.  Unless the
mass scale $b$ of the tensor field is close to the Planck scale, we
will have $\epsilon \ll 1$, and the growth of $g_{tt}$ should be
sufficiently slow that we can ``patch'' our solution into one
describing the appropriate larger-scale structure.  Thus, although
this solution behaves oddly in the asymptotic region, its behaviour
does not seem bad enough to reject it as unphysical.

With this asymptotic geometry found, we can now ask what its effects
on the propagation of test particles (particularly light rays) might
be.  Assuming that our tensor field $B_{ab}$ does not couple directly
to the Maxwell field, a light ray propagating in this background will
follow a null geodesic.  Without loss of generality, we can assume
that this geodesic lies in the equatorial plane (i.e., $\theta =
\pi/2$.)  Since our spacetime is static and spherically symmetric, it
has two Killing vector fields $t^a$ and $\phi^a$ (the generators of
timelike translations and rotations in the equatorial plane,
respectively) giving rise to two constants of the motion:
\begin{align}
  \label{geoconst}
  E &= -u^a t_a, & L &= u^a \phi_a,
\end{align}
where $u^a$ is the four-velocity of the particle.

Two possible physical effects spring to mind that might arise in a
geometry such as this one: gravitational redshift and deflection of
light rays.  Using standard techniques \cite{Wald}, it can be shown
that a light ray emitted with frequency $\omega_e$ at some distance $r_e$
from the monopole will be observed to have a frequency $\omega_o$ by an
observer at a distance $r_o$ from the monopole, where
\begin{equation}
  \label{redshift}
  \frac{\omega_o}{\omega_e} = \frac{M(r_e)}{M(r_o)} = \left(
    \frac{r_e}{r_o} \right)^\epsilon.
\end{equation}
If $\epsilon \ll 1$, the first-order fractional redshift $\Delta
\omega/\omega_e$ will then be
\begin{equation}
  \label{fracred}
  \frac{\Delta \omega}{\omega_e} \approx \epsilon \ln
  \left(\frac{r_e}{r_o} \right).
\end{equation}
We can see that any gravitational redshift due to the presence of the
monopole will be quite small, especially if $r_e$ and $r_o$ are close
to the same order of magnitude.  Even if $r_e$ and $r_o$ are
unrealistically disparate in size---say, the Planck distance and the
Hubble distance, respectively---we would still have $\ln (r_e/r_o)
\approx 140$, and a mass scale $b$ that was more than a few orders of
magnitude less than the Planck scale would still make this redshift
extremely difficult to detect.

The situation for the deflection of light by the spacetime curvature
is somewhat more interesting.  We again use the standard techniques
\cite{Wald} to derive the motion of massless particles.  A null
geodesic in the equatorial plane will satisfy
\begin{equation}
  - M^2(r) \dot{t}^2 + N^2(r) \dot{r}^2 + r^2 \dot{\phi}^2 = 0,
\end{equation}
where a dot over a symbol denotes the derivative of the particle's
coordinate position $(t(s), r(s), \pi/2, \phi(s))$ with respect to
some affine parameter $s$ on the worldline.  The constants of the
motion \eqref{geoconst} are given by $E = M^2 \dot{t}$ and $L = r^2
\dot{\phi}$; we can thus write
\begin{equation}
  \dot{r}^2 + \left[ \frac{L^2}{N^2 r^2} -
    \frac{E^2}{M^2 N^2} \right] = 0. 
\end{equation}
Using the asymptotic forms of $M$ and $N$ found above (with $C_1 \to
0$), we see that the path of the null geodesic satisfies
\begin{equation}
  \label{phirpath}
  \frac{d \phi}{d r} = \frac{\dot{\phi}}{\dot{r}} = \pm \frac{\sqrt{1 +
    \epsilon}}{r^2 \sqrt{ \beta^{-2} (\sqrt{2 \lambda} b)^{-2 \epsilon}
    r^{-2 \epsilon} - r^{-2} }}.
\end{equation}
where $\beta \equiv L/E$.\footnote{In an asymptotically flat
  spacetime, $\beta$ would be the ``apparent impact parameter'' of the
  light ray.  In our situation, it is not immediately clear what the
  physical interpretation of this parameter is; mathematically,
  however, it plays much the same role.}  We can then find the total
angular deflection of this geodesic by integrating this quantity over
$r$ from $\infty$ to $r_m$ (the value of $r$ for which the denominator
of \eqref{phirpath} vanishes) and doubling it (to account for both the
deflection incurred travelling from $r = \infty$ to $r = r_m$ and the
deflection incurred in travelling back to $r = \infty$.)  The result
is\footnote{This integral can be done by switching to a coordinate $u
  \equiv (r_m/r)^{2-2\epsilon}$; the resulting integral can then be
  seen to be proportional to the Euler beta function.}
\begin{equation}
  \label{defangle}
  \Delta \phi = \pi \frac{\sqrt{1 + \epsilon}}{1 - \epsilon}, 
\end{equation}
or, defining $\delta \phi \equiv \Delta \phi - \pi$ to be the angle
between the ``unperturbed'' and ``perturbed'' directions of
propagation,
\begin{equation}
  \delta \phi \approx \frac{3 \pi}{2} \epsilon.
\end{equation}

We see that to leading order in $\epsilon$, the deflection angle
$\delta \phi$ does not depend on $\beta$, and thus is independent of
the properties of the geodesic.  In other words, with respect to the
propagation of light, at lowest order this spacetime behaves as though
it has a solid deficit angle but is otherwise flat.  This ``apparent
deficit angle'' is not the same as the deficit angle of the
constant-time slices mentioned above; it arises from both the deficit
angle in the spatial geometry and from the behaviour of the
$tt$-component of the metric (roughly analogous to the gravitational
potential.)  From an observational point of view, however, the
light-bending signature of one of our antisymmetric tensor monopoles
would be exactly the same as that of the previously examined $O(3)$
scalar monopoles \cite{BarrVil}; only the dependence of the deflection
angle on the respective mass scales of the two models differs.

\section{Discussion}
\label{discsec}

We have examined the existence and properties of topological defect
solutions arising from a spacetime tensor which spontaneously breaks
Lorentz symmetry by acquiring a fixed norm $\mathcal{T}_{(a)}
\mathcal{T}^{(a)} = C$ in vacuum.  For topological defect solutions to
exist, the set of such tensor fields (the vacuum manifold) must
contain a non-contractible $S^0$, $S^1$, or $S^2$; we found that the
only tensors with rank not greater than five and of definite symmetry
type with this topology are vectors, antisymmetric two-tensors, and
symmetric two-tensors.  From these, we obtained domain wall solutions
in which a vector takes on a negative (timelike) norm asymptotically,
and monopole solutions in which an antisymmetric two-tensor takes on a
positive norm asymptotically.  These vector domain wall solutions
appear to be unstable; however, the antisymmetric two-tensor monopole
solutions are well-behaved in flat spacetime, and can give rise to
observable light-bending effects.

It is notable that we have not found any cosmic string solutions in
our work.  Such solutions would require the signature of our tensor
space to have either $n_+ = 2$ or $n_- = 2$; consulting Table
\ref{sigtable}, we see that no such tensor with $r \leq 5$ exists.  It
is unclear whether a deep mathematical reason exists for this lacuna;
such a tensor space might exist at higher rank, but this seems
unlikely.

If cosmic string solutions are desired, two approaches are still open.
The first is simply to combine multiple fields.  Rather than having a
single tensor field $\mathcal{T}^{(a)}$ with a potential minimized at
$\mathcal{T}_{(a)} \mathcal{T}^{(a)} = C$, we can envision two tensor
fields $(\mathcal{T}_1)^{(a)}$ and $(\mathcal{T}_2)^{(a)}$ with a
combined potential that is minimized when
\begin{equation}
  \label{combo}
  (\mathcal{T}_1)^{(a)}(\mathcal{T}_1)_{(a)} \pm
  (\mathcal{T}_2)^{(a)}(\mathcal{T}_2)_{(a)} = C.
\end{equation}
If the respective signatures of the tensor spaces in which
$(\mathcal{T}_1)^{(a)}$ and $(\mathcal{T}_2)^{(a)}$ lie are $(n_{1+},
n_{1-})$ and $(n_{2+},n_{2-})$, respectively, then it is not hard to
see that the ``effective'' signature of the combined tensor space is
$(n_{1+} + n_{2\pm}, n_{1-} + n_{1\mp})$.  Thus, by combining two or
more fields such that some linear combination of their norms is
minimized, we could in principle obtain a vacuum manifold with an
``effective'' $n_+$ or $n_-$ equal to two.  

Unfortunately, an examination of Table \ref{sigtable} shows that the
only two types of tensor fields with $n_+ \leq 2$ or $n_- \leq 2$ are
scalar and vector fields (or fields equivalent to these.)  The case of
two scalar fields (or a single complex scalar field) is already
well-known, and does not violate Lorentz symmetry in the sense we are
interested in.  A theory containing a scalar field and a vector field,
in a potential that is minimized when $A^a A_a - \phi^2 = -b^2$, would
have a non-contractible $S^1$ in its vacuum manifold; so would a
theory of a complex vector field (effectively, two independent vector
fields) with $A^a A^*_a = - b^2$.  However, since we would still be
using the timelike components of the vector fields to construct our
topological defect, we would again need to ``invert'' the potential
(as in the case of a single vector field) to obtain a global solution.
Such theories would likely then share the same instability that our
original vector domain-wall model had.

The other option to obtain cosmic string solutions would be to
generalize our vacuum manifolds.  Throughout this work, we have been
examining only those sets of tensors for which the ``square'' of the
tensor is some constant value.  It is plausible that some other
invariant (of higher order than quadratic in the tensor field) could
give a vacuum manifold of the correct topology to support cosmic
string defects.  Additionally, more complicated potentials could give
rise to defect types that would be impossible via a simple ``sum of
norms'' vacuum manifold of the form \eqref{combo}; an example of this,
in which two spacelike vector fields form a domain wall solution, was
described in \cite{fancyvecwall}.  Such manifolds could be expressed
as the set of zeroes of one or more polynomials in $\mathbb{R}^n$ for
some $n$, also known as a real algebraic variety.  Unfortunately, it
does not appear that any known characterization of the topology of
such spaces is as complete as that for the quadratic case; see
\cite{BCR} for more details.

Both of the solutions we have found were obtained under the assumption
of the maximal symmetry compatible with the type of defect solution we
sought: planar symmetry in the domain wall case, spherical symmetry in
the monopole case, and staticity in both cases.  It is likely that
that interesting solutions with a lesser degree of symmetry might also
exist in these theories, especially if one relaxes the staticity
requirement; one could look at linearized solutions about these
backgrounds and investigate their evolution.  (As noted above, the
evolution of perturbations of a domain wall is likely to be
unbounded.)  It is also conceivable that solutions with reduced
spatial symmetry but which are still static might also exist, both for
these theories and for theories containing symmetric tensors and
spacelike vector fields.  (We previously rejected these latter two
types of fields as uninteresting due to their lack of symmetric
solutions.)  Intuitively, such solutions would seem less likely to
exist, and would most likely be of higher energy than the symmetric
solutions of the theory; certainly, such solutions would not be
``close'' in any real sense to the symmetric solutions we have found.

In previous work on topological defects, it has been useful to draw a
distinction between ``global'' monopoles, in which a global symmetry
is broken, and ``gauge'' monopoles, in which a local symmetry is
broken.  In the present case, it is unclear how useful this
distinction is.  In the case of a fixed flat background, the Lorentz
and Poincar\'e symmetries of the Lagrangian are global ones, so in
this case we can classify our solutions as global monopoles.  Since
diffeomorphism symmetry can be thought of as ``gauged Poincar\'e
symmetry'' \cite{KibbleLor}, one could then say that a gauged solution
is one where the Levi-Citiva connection and (by extension) the metric
are dynamical fields; from this perspective, a gravitating monopole is
a gauge monopole.  However, unlike in the case of scalar monopoles,
the passage from global symmetry to gauge symmetry does not greatly
change the behaviour of the original global monopole solutions.  In
particular, the $O(3)$ scalar monopoles have an energy (i.e., the
integral of the energy density over all of space) that is formally
infinite, due to the $r^{-2}$ fall-off rate of the energy density
$\rho$.  This infinity is eliminated when this $O(3)$ symmetry is
promoted to a gauge symmetry, as the parts of the fields that cause
the divergence can in effect be ``gauged away''.  In contrast, our
antisymmetric tensor monopoles retain the same energy density fall-off
properties when we promote the metric to a dynamical field, and so are
not cured of their formally divergent energies.  It therefore seems
that the distinction between global and gauge monopoles is not as
physically relevant in the present case as it is in the case of
internal symmetries.

The experimental prospects for the observation of these antisymmetric
tensor monopoles will, of course, depend critically on their current
abundance in the Universe.  We would expect that such topological
defects would have formed in a phase transition as the Universe cooled
after the Big Bang, via the Kibble mechanism \cite{Kibble}.  To within
an order of magnitude, one such structure should form in each Hubble
volume at the time of this phase transition.  However, it is not clear
how efficiently these structures might recombine in the subsequent
evolution of the Universe.  Much of Barriola \& Vilenkin's discussion
\cite{BarrVil} concerning the recombination of global $O(3)$ monopoles
applies here.  Since these are global monopoles, the characteristic
energy scale of a monopole-antimonopole pair will be directly
proportional to the distance separating them; the effective force
between them will then be independent of distance.  This would seem to
imply that pair-annihilation of such structures would be quite
efficient.  It is unclear, however, how easily these monopoles can
``find each other'' in an expanding Universe.  As in the $O(3)$ case,
a numerical simulation will probably be required to answer the
question of current abundance of AS tensor monopoles.  Simulations in
the $O(3)$ case \cite{BennRhie} have shown that the density of such
monopoles remains roughly constant at approximately $(4.0 \pm 1.5)
d_H^{-3}$, where $d_H$ is the horizon distance, as the Universe
evolves; it seems plausible that similar results (up to an order of
magnitude) will obtain in our case.

\acknowledgments

I would like to thank B.\ Altschul, C.\ Deffayet, D.\ Garfinkle, V.\
A.\ Kosteleck\'y, and M.\ Uhlmann for important ideas and discussion
concerning this work.  This work was supported in part by the United
States Department of Energy under Grant No.\ DE-FG02-91ER40661.

\appendix*

\section{Derivation of tensor space signatures}

In Table \ref{sigtable}, we gave a list of the signatures $(n_+,n_-)$
of tensor spaces with rank $r \leq 5$ and definite symmetry pattern.
We now present the method by which these signatures were found.  Much
of what follows, particularly in the first subsection, is based on the
classic texts by Hamermesh \cite{Hamermesh} and by Weyl \cite{Weyl}.

\subsection{Preliminaries}
\label{sigprelim}

\subsubsection{Representations of $GL(d)$}

Let $V$ denote a $d$-dimensional real vector space, and let
$\mathbb{U}_r$ denote the tensor product of $V$ with itself $r$ times;
in other words, the elements of $\mathbb{U}_r$ are the rank-$r$
contravariant tensors on $V$.  This space is itself a
$d^r$-dimensional vector space: it is closed under addition and
multiplication by real numbers.

Consider now the group of all non-singular linear transformations on
$V$, denoted by $GL(d)$.  The action of this group on $V$ extends in a
natural way to the tensor space $\mathbb{U}_r$.  (Roughly speaking,
given an element $g \in GL(d)$, we can act on each ``copy'' of $V$ in
$\mathbb{U}_r$ with $g$.)  This group action can be seen to be a
faithful representation of $GL(d)$ with $\mathbb{U}_r$ as its
representation space.  It is further known that this representation is
reducible, i.e., the space $\mathbb{U}_r$ can be written as
\begin{equation}
  \mathbb{U}_r = \bigoplus_i \mathbb{V}_i,
\end{equation}
such that each subspace $\mathbb{V}_i$ is closed under the above
described action of $GL(d)$.  

These subspaces $\mathbb{V}_i$ are essentially obtained by resolving
an arbitrary tensor into tensors with some symmetry among their
indices.  A familiar example of this is the case $r=2$: an arbitrary
tensor $t^{ab}$ can be resolved into two parts, one symmetric and one
antisymmetric, i.e.,
\begin{equation}
  t^{ab} = t^{(ab)} + t^{[ab]}.
\end{equation}
We can write this in terms of projectors on the space of tensors;
defining
\begin{equation}
  (P_S)^{a_1 a_2} {}_{b_1 b_2} = \frac{1}{2} (\eta^{a_1} {}_{b_1} 
  \eta^{a_2} {}_{b_2} + \eta^{a_1} {}_{b_2} 
  \eta^{a_2} {}_{b_1})
  \label{symex}
\end{equation}
and
\begin{equation}
  (P_A)^{a_1 a_2} {}_{b_1 b_2} = \frac{1}{2} (\eta^{a_1} {}_{b_1} 
  \eta^{a_2} {}_{b_2} - \eta^{a_1} {}_{b_2} 
  \eta^{a_2} {}_{b_1}),
  \label{asymex}
\end{equation}
it is not hard to show that 
\begin{subequations}
\label{proj2}
\begin{equation}
    (P_S)^{a_1 a_2} {}_{b_1 b_2} (P_S)^{b_1 b_2} {}_{c_1 c_2} =
    (P_S)^{a_1 a_2} {}_{c_1 c_2},
\end{equation}
\begin{equation}
    (P_A)^{a_1 a_2} {}_{b_1 b_2} (P_A)^{b_1 b_2} {}_{c_1 c_2} =
    (P_A)^{a_1 a_2} {}_{c_1 c_2}, 
\end{equation}
\begin{equation}
    (P_S)^{a_1 a_2} {}_{b_1 b_2} (P_A)^{b_1 b_2} {}_{c_1 c_2} =
    (P_A)^{a_1 a_2} {}_{b_1 b_2} (P_S)^{b_1 b_2} {}_{c_1 c_2} =
    0,
\end{equation}
\end{subequations}
and
\begin{equation}
  \label{identresol2}
  (P_S)^{a_1 a_2} {}_{b_1 b_2} + (P_A)^{a_1 a_2} {}_{b_1 b_2} =
  \eta^{a_1} {}_{b_1} \eta^{a_2} {}_{b_2}.
\end{equation}
Thus, $(P_S)^{a_1 a_2} {}_{b_1 b_2}$ and $(P_A)^{a_1 a_2} {}_{b_1
  b_2}$ are projectors in the $d^2$-dimensional space of two-tensors
on $V$. $(P_S)^{(a)} {}_{(b)}$ projects an arbitrary tensor $t^{ab}$
to its symmetric part, while $(P_A)^{(a)} {}_{(b)}$ projects $t^{ab}$
onto its antisymmetric part.  We can denote the subspaces projected
onto by $(P_S)^{(a)} {}_{(b)}$ and $(P_A)^{(a)} {}_{(b)}$ as
$\mathbb{V}_S$ and $\mathbb{V}_A$, respectively.  Moreover, the action
of $GL(d)$ will map symmetric tensors to symmetric tensors and
antisymmetric tensors to antisymmetric tensors; thus, $\mathbb{V}_S$
and $\mathbb{V}_A$ are closed under the action of $GL(d)$.

To generalize this to higher-rank tensors, we must first introduce an
algebra acting on $\mathbb{U}_r$.  Let $S_r$ be the group of
permutations on $r$ objects, and let $\mathfrak{s}_r$ be the
$r!$-dimensional vector space spanned by tensors of the form
\begin{equation}
  \label{etaperm}
  \eta^{a_{\sigma(1)}} {}_{b_1} \eta^{a_{\sigma(2)}} {}_{b_2} \cdots
      \eta^{a_{\sigma(r)}} {}_{b_r}  \equiv \eta^{(a_{\sigma(i)})}
        {}_{(b_i)}
\end{equation}
for some permutation $\sigma \in S_r$.  The action of
$\eta^{(a_{\sigma(i)})} {}_{(b_i)}$, when contracted with a tensor
$F^{(b_i)}$, is simply to return that tensor with its indices
rearranged by the permutation $\sigma$. By definition, we can add two
elements of $\mathfrak{s}_r$ together to form another element of
$\mathfrak{s}_r$, or multiply any element by a real number.  However,
we can also multiply elements of $\mathfrak{s}_r$ together in a
natural way: if $a = \eta^{(a_{\sigma(i)})} {}_{(b_i)}$ is a basis
element corresponding to a permutation $\sigma \in S_r$ and $b =
\eta^{(a_{\tau(i)})} {}_{(b_i)}$ corresponds to $\tau \in S_r$, then
we can define the product of $a$ and $b$ as
\begin{equation}
  ab = \eta^{(a_{\sigma(i)})} {}_{(b_i)} \eta^{(b_{\tau(i)})}
  {}_{(c_i)} = \eta^{(a_{\sigma(\tau(i))})} {}_{(c_i)},
\end{equation}
since $\eta^{(a_{\sigma(i)})} {}_{(b_i)} =
\eta^{(a_{\sigma(\tau(i))})} {}_{(b_{\tau(i)})}$.  (This can be seen
by an appropriate rearrangement of the $\eta$'s in the definition
\eqref{etaperm} above.)  When viewed in terms of its action of
$\mathbb{U}_r$, this implies that acting on $F^{(a)}$ by the element of
$\mathfrak{s}$ associated with $\tau$, followed by that element
associated with $\sigma$, yields $F^{(a)}$ with its indices permuted
by $\sigma \tau$, exactly as one would expect.  The multiplication of
two arbitrary vectors in $\mathfrak{s}_r$ is then given by requiring
distributivity to hold for our multiplication operator, i.e., $(a +
b)c = ac + bc$ and $c(a+b) = ca + cb$ for all $a$, $b$, and $c$ in
$\mathfrak{s}_r$.

Not all elements in $\mathfrak{s}_r$ are associated with strict
permutations; a general element of $\mathfrak{s}_r$, when contracted
with a tensor $F^{(a)}$, will return some linear combination of the
various index permutations of $F^{(a)}$.  Such linear combinations are
what are required to describe the decomposition of $\mathbb{U}_r$ into
irreducible subspaces under $GL(d)$.  In the $r=2$ example above, we
found two elements of $\mathfrak{s}_2$, $(P_S)^{(a)} {}_{(b)}$ and
$(P_A)^{(a)} {}_{(b)}$, which acted as projectors on the space
$\mathbb{U}_2$ \eqref{proj2}; moreover, these projectors resolved the
identity \eqref{identresol2}.  In general, it can be shown that such a
set of projectors $Y_i \in \mathfrak{s}_r$, called \emph{Young
  symmetrizers}, exists for arbitrary $r$; these symmetrizers can be
constructed by means of Young tableaux, and each symmetrizer has
associated with it a particular Young tableau.  For further details,
the interested reader is referred to Hamermesh \cite{Hamermesh}.

\subsubsection{$GL(d) \to GL(d-1)$ decomposition}
\label{GLtoGLthy}

The signature of the metric $\eta_{(a)(b)}$ on an irreducible $GL(d)$
subspace $\mathbb{V}$ is, of course, an indefinite metric; roughly
speaking, every ``time'' component of a tensor gives a negative sign,
while the ``space'' components give positive signs.  It will therefore
be advantageous to look at how a given tensor representation behaves
under purely ``spatial'' transformations, i.e., those which act on the
some set of positive-norm coordinates $\{x^1, x^2, \dots, x^{d-1} \}$
while leaving the negative-norm ``time'' coordinate $x^d$
invariant.\footnote{Throughout this appendix, we will take $x^d$ to be
  our time coordinate.}

Under this subgroup $GL(d-1) \subset GL(d)$, each irreducible subspace
$\mathbb{V}$ will split into the direct sum of several subspaces
$\mathbb{W}_i$, each of which is invariant under $GL(d-1)$:
\begin{equation}
  \mathbb{V} = \bigoplus_i \mathbb{W}_i.
  \label{gld1decomp}
\end{equation}
These subspaces are themselves equivalent (in terms of their behaviour
under $GL(d-1)$) to tensor spaces of definite symmetry type;
specifically, the Young tableaux of the subspaces can be obtained by
removing one box from the bottom of some or all of the columns of the
Young tableau corresponding to $\mathbb{V}$, making sure that the
resultant pattern is in fact a valid tableau.  For example, we have
\begin{equation}
\yng(2,2) \to \yng(2,2) \oplus \yng(2,1) \oplus \yng(2).
\end{equation}

To see this, consider the ``standard components'' of a tensor in
$\mathbb{V}$, i.e., a complete set of components which determine all
the others via the symmetry relations on $\mathbb{V}$.  If
$\mathbb{V}$ is the image of some Young symmetrizer $Y$, the standard
components of a tensor in $\mathbb{V}$ are found by taking the
standard tableau corresponding to $Y$ and filling with the symbols
$\{1, \dots, d \}$ such that the symbols are non-decreasing along the
rows and strictly increasing down the columns.\footnote{For example,
  under the Young symmetrizer corresponding to {\scriptsize
    \young(\ai\aiii,\aii)}, the filling {\scriptsize \young(11,2)}
  corresponds to the $F_{121}$ component of the tensor $F_{a_1 a_2
    a_3}$; the filling {\scriptsize \young(12,3)} corresponds to the
  $F_{132}$ component, etc.}  Importantly, this means that the symbol
$d$ will only appear at the bottom of a column in such a filling.  We
can then consider the subspace of $\mathbb{V}$ spanned by the set of
tensors $\{ (e_i)^{(a)} \}_{\mathbb{V}}$, where the non-vanishing
standard components of each tensor $(e_i)^{(a)}$ are those with $d$'s
at the bottom of a certain subset of the columns in the Young tableau
for $\mathbb{V}$.  Since the action of $GL(d-1)$ effectively ``leaves
$d$'s alone'', such a subspace will be invariant under the action of
$GL(d-1)$.

\subsubsection{$GL(d) \to O(d-1,1)$ decomposition}
\label{GLtoOthy}

We have thus far been examining the properties of tensors in
$\mathbb{U}_r$ under arbitrary invertible transformations on the
underlying vector space $V$.  However, this symmetry group is not the
physically relevant one.  Rather, we expect our physical laws to be
invariant under the actions of the Lorentz subgroup $O(d-1,1) \subset
GL(d)$, defined as that subgroup of $GL(d)$ which leaves the spacetime
metric $\eta_{ab}$ (and, by extension, the tensor metric
$\eta_{(a)(b)}$) invariant.  As in the $GL(d-1) \subset GL(d)$ case,
an irreducible space $\mathbb{V}$ will decompose into another set of
invariant subspaces $\bar{\mathbb{W}}_i$:
\begin{equation}
  \label{sodecomp}
  \mathbb{V} = \bigoplus_i \bar{\mathbb{W}}_i.
\end{equation}
(These spaces $\bar{\mathbb{W}}_i$ will not generically be the same as
the $GL(d-1)$ irreducible spaces $\mathbb{W}_i$.)  Without loss of
generality, let the Young symmetrizer $Y$ for $\mathbb{V}$ be obtained
from a standard filling of a Young tableau in which as many of the
pairs $\{ \{a_1, a_2\}$, $\{a_3, a_4\}$, $\{a_5, a_6\} \dots \}$ as
possible are in the same row; in other words, the resultant tensors
will be symmetric under the exchange of $a_1$ and $a_2$, of $a_3$ and
$a_4$, and so forth.  In general, we can then decompose an arbitrary
tensor $F^{(a)} \in \mathbb{V}$ into its components in invariant
subspaces under $\bar{\mathbb{W}}_i$ by applying the Young symmetrizer
to products of the inverse metric $\eta^{ab}$ and \emph{trace-free}
tensors $(f_i)$ of rank $r, r-2, r-4, \dots$:
\begin{multline}
  F^{(a)} = f_1^{(a)} + Y^{(a)} {}_{(b)} \left[ \sum_i
    \eta^{b_1 b_2} (f_i)^{b_3 \dots 
      b_r} \right] \\ + Y^{(a)} {}_{(b)} \left[ \sum_i \eta^{b_1 b_2}
    \eta^{b_3 b_4} (f_i)^{b_5 \dots b_r}\right] + 
  \dots. 
    \label{tracefreedecomp}
\end{multline}
Moreover, these tensors $(f_i)$ can each be taken be of definite
symmetry type, with a Young pattern of their own; in other words, they
lie within some irreducible $GL(d)$ component of $\mathbb{U}_{r-2t}$
for some $t$.  

In general, a Ferrers diagram\footnote{A Ferrers diagram is
  essentially an unfilled Young tableau.  We will denote such diagrams
  by a list of numbers indicating the lengths of their rows; for
  example, $\{ 3,1 \}$ denotes the diagram {\scriptsize \yng(3,1)}. }
$\{ \lambda'_1, \dots, \lambda'_{m'} \}$ will appear on the right-hand
side of the $GL(d) \to O(d-1,1)$ decomposition of an arbitrary Ferrers
diagram $\{\lambda_1, \dots, \lambda_m \}$ if $\{\lambda_1, \dots,
\lambda_m \}$ is contained in the tensor product of $\{ \lambda'_1,
\dots, \lambda'_{m'} \}$ and $\{ \delta_1, \dots, \delta_{m''} \}$,
\emph{where all the integers $\delta_i$ are even.}  If it is possible
to multiply $N > 1$ such diagrams $\{ \delta_1, \dots, \delta_{m''}
\}$ with $\{ \lambda'_1, \dots, \lambda'_{m'} \}$ to obtain
$\{\lambda_1, \dots, \lambda_m \}$, then there will be $N$ distinct
subspaces with the Ferrers diagram $\{ \lambda'_1, \dots,
\lambda'_{m'} \}$ in the decomposition.  However, this only occurs for
$r \geq 6$;\footnote{Specifically, the representation given by the
  Ferrers diagram $\{\lambda_i \} = \{4,2 \}$ can be obtained by
  multiplying $\{ \lambda'_i \} = \{2\}$ by either $\{\delta_i\} =
  \{4\}$ or $\{ \delta_i \} = \{2,2\}$; thus, the decomposition of the
  space of tensors of type $\{4,2\}$ under $GL(d) \to O(d-1,1)$ will
  contain two subspaces equivalent to the space of trace-free
  symmetric tensors (those of type $\{2\}$.) \label{rank6bad}} for $r
\leq 5$, the invariant subspaces are uniquely labelled by their
symmetry types. See \cite{Littlewood, Jahn} for further details.

As an example, if $\mathbb{V}$ is the space of all tensors with
symmetry type given by the Ferrers diagram $\{3,1\}$, it can be shown
that\begin{equation} \yng(3,1) \to \yng(3,1) \oplus \yng(2) \oplus
  \yng(1,1).
\end{equation}
In other words, $\mathbb{V}$ decomposes into three invariant subspaces
under $O(d-1,1)$: the space of all trace-free tensors in $\mathbb{V}$,
a space equivalent under $O(d-1,1)$ to the space of all trace-free
symmetric two-tensors, and a space equivalent to the space of all
antisymmetric two-tensors.  The decomposition \eqref{tracefreedecomp}
of an arbitrary tensor in $F^{(a)} \in \mathbb{V}$ would then have
three terms on the right-hand side, one corresponding to each of its
components in each subspace $\bar{\mathbb{W}}_i$.  One of these terms
would be the trace-free rank-4 tensor $f_1^{a_1 a_2 a_3 a_4}$.  The
other two would come from the first sum in \eqref{tracefreedecomp};
they would each be expressible as the product of $\eta^{a_1 a_2}$ with
a rank-two tensor ($f_2^{a_3 a_4}$, a symmetric trace-free tensor, and
$f_3^{a_3 a_4}$, an antisymmetric tensor, respectively) and
appropriately symmetrized by $Y^{(a)} {}_{(b)}$.

\subsection{Signatures of tensor spaces}

\subsubsection{General tensor spaces}

As noted in Section \ref{vacmansec} above, the ``tensor norm'' given
in \eqref{tensornorm} can be thought of as a quadratic form on
$\mathbb{U}_r$ or any subspace thereof.  If we are dealing with the
entire space $\mathbb{U}_r$ of rank-$r$ tensors, we can easily
construct a basis for this space; simply pick orthonormal coordinates
$\{ x^1, x^2, \dots x^d \}$ on our space $V$.  We can
then construct a set of $d^r$ tensors $\{(e_i)^{(a)}\} \subset
\mathbb{U}_r$ for each one of which a single coordinate component
$F^{\mu_1 \mu_2 \cdots \mu_r}$ is set to unity and the rest vanish.
These tensors can be seen to be an orthonormal basis for
$\mathbb{U}_r$ under the quadratic form induced by $\eta_{(a)(b)}$.
Moreover, since the signature of our spacetime $(n_+, n_-)$ is
$(d-1,1)$, it is not hard to describe which basis elements have
positive norm and which have negative norm.  If the non-vanishing
component $F^{(\mu)}$ of one of our basis tensors has an even number
of ``time indices'' (i.e., if an even number of the $\mu_i$ equal $d$),
then the norm of this tensor will contain an even number of factors of
$\eta_{dd}$ and will be positive.  Similarly, if the non-vanishing
component has an odd number of time indices, the norm of this tensor
will be negative.  Some combinatorics then shows that the number of
basis tensors with positive norm is
\begin{equation}
  n_+ = \frac{1}{2} (d^r + (d-2)^r),
\end{equation}
while the number of basis tensors with negative norm is
\begin{equation}
  n_- = \frac{1}{2} (d^r - (d-2)^r).
\end{equation}
For $d = 4$, the signatures $(n_+,n_-)$ for $r = \{0,1,2,3,\dots\}$
will be $\{(1,0), (3,1), (10,6), (36,28), \dots \}$.

We have thus found the signature $(n_+,n_-)$ of the space
$\mathbb{U}_r$ of all tensors of rank $r$.  However, we noted in the
previous subsection that the space $\mathbb{U}_r$ decomposes into the
direct sum of several spaces $\mathbb{V}_i$ of fixed symmetry type.
It is natural then to ask what the signature of the metric
$\eta_{(a)(b)}$, confined to one of these subspaces (call it
$\mathbb{V}$), is.  However, it is not immediately obvious how to
construct a basis for $\mathbb{V}$ that is orthonormal under
$\eta_{(a)(b)}$, or even that such a basis exists (for all we know,
$\eta_{(a)(b)}$ might be degenerate on $\mathbb{V}$.)

A clue as to how to proceed can be gleaned from our treatment of the
signature of $\mathbb{U}_r$ above.  Consider subspace of
$\mathbb{U}_r$ spanned by the subset of all of the basis tensors $\{
(e_i)^{(a)} \}$ whose non-vanishing component has a
certain subset of the indices $\mu_i$ equal to $d$ and the rest
differing from $d$.  (For example, for an arbitrary two-tensor
$t^{ab}$, we might consider those basis elements for which $t^{1d},
t^{2d}, \dots t^{(d-1)d}$ are non-vanishing; $t^{dd}$ and $t^{d1}$
would be in the spanning set of different such subspaces.)  While
these subspaces (call them $\bar{\mathbb{U}}_i$) are not invariant
under an arbitrary linear transformation in $GL(d)$, the subgroup
$GL(d-1)$ which leaves the timelike vector $(0,0,\dots,0,1)$ invariant
will also leave each $\bar{\mathbb{U}}_i$ invariant.  Moreover, we can
see that the quadratic form induced by $\eta_{(a)(b)}$ is either
positive or negative definite on each such subspace, and that the sign
of $\eta_{(a)(b)}$ on each subspace is determined by the number of
``time components'' corresponding to each subspace.

We wish to find a similar procedure for an invariant $GL(d)$ subspace
$\mathbb{V}$, by using the $GL(d) \to GL(d-1)$ decomposition described
in Section \ref{GLtoGLthy}.  The first complication arises when we try
to determine the orthogonality of the irreducible subspaces
$\mathbb{W}_i$.  In the case of the above decomposition of
$\mathbb{U}_r$, it was fairly evident that the irreducible $GL(d-1)$
subspaces $\bar{\mathbb{U}}_i$ were orthogonal; thus, the union of the
orthonormal bases for these subspaces formed an orthonormal basis for
$\mathbb{U}_r$, and we could read off $\mathbb{U}_r$'s signature by
knowing the sign of the metric $\eta_{(a)(b)}$ on the several
$\bar{\mathbb{U}}_i$ along with their respective dimensionalities.  In
the case of $\mathbb{V}$, however, things are not so clear.  It is
fairly evident that any two subspaces $\mathbb{W}_i$ whose Young
tableaux are obtained from the Young tableau of $\mathbb{V}$ via the
removal of differing numbers of boxes will be orthogonal; when we take
the inner product between two tensors $F_1^{(a)}$ and $F_2^{(a)}$ lying in
two such spaces,
\begin{equation}
  F_1^{(a)} F_2^{(b)} \eta_{(a)(b)} = F_1^{a_1 \cdots a_r} F_2^{b_1 \cdots
    b_r} \eta_{a_1 b_1} \dots \eta_{a_r b_r},
\end{equation}
at least one of the $\eta_{a_i b_i}$'s will have one $d$ index and one
non-$d$ index, and so the whole thing will vanish.  However, this
argument does not work in the case of two spaces $\mathbb{W}_i$ whose
respective tableaux are obtained by removal of the same number of
boxes;  as an example, consider the example
\begin{equation}
  \yng(3,1) \to \yng(3,1) \oplus \yng(3) \oplus \yng(2,1) \oplus \dots.
\end{equation}
It is not immediately clear that the inner product of two tensors
lying in each of the last two subspaces above will be zero.  To show
that two such spaces are orthogonal requires other techniques;
specifically, it follows from the following theorem:
\begin{theorem}
  \label{orthirreds}
  Let $\mathbb{V}_1$ and $\mathbb{V}_2$ be two irreducible subspaces
  of $\mathbb{U}_r$ whose corresponding Young symmetrizers $Y_1$ and
  $Y_2$ correspond to tableaux of differing shape.  Let $F_1^{(a)} \in
  \mathbb{V}_1$ and $F_2^{(a)} \in \mathbb{V}_2$. Let $P \in
  \mathfrak{s}_r$.  Then $\eta_{(a)(b)} F_1^{(a)} P^{(b)} {}_{(c)}
  F_2^{(c)} = 0$.
\end{theorem}
If this theorem holds, then it is not hard to see that our subspaces
$\mathbb{W}_i$ must all be orthogonal.  Denote by $V'$ the subspace of
$V$ consisting of all vectors $v^a$ with the component $v^d = 0$, and
denote by $\mathbb{U}'_r$ the tensor product of $V'$ with itself $r$
times.  Let $\mathbb{W}_1$ and $\mathbb{W}_2$ be two irreducible
subspaces of $\mathbb{V} \subset \mathbb{U}_r$, which both transform
as rank $r'$ tensors under the action of $GL(d-1) \subset GL(d)$.
This means that there exists a linear bijection $\phi_1$ ($\phi_2$)
mapping $\mathbb{W}_1$ ($\mathbb{W}_2$) into $\mathbb{V}'_1$
($\mathbb{V}'_2$), an irreducible subspace of $\mathbb{U}'_{r'}$ with
Young symmetrizer $Y_1$ ($Y_2$).  When we take the inner product of
two tensors $F_1^{(a)} \in \mathbb{W}_1$ and $F_2^{(a)} \in
\mathbb{W}_2$, this will correspond to some complete contraction
(possibly with indices permuted, and possibly a sum of such terms)
between the corresponding tensors $f^{(\tilde{a})} =
\phi_1(F_1^{(a)})$ and $f_2^{(\tilde{a})} = \phi_2(F_2^{(a)})$.  In
other words, there will exist a $P \in \mathfrak{s}_{r'}$ such that
\begin{equation}
F_1^{(a)} F_2^{(b)} \eta_{(a)(b)} = {\eta'}_{(\tilde{a})(\tilde{b})}
f_1^{(\tilde{a})} P^{(\tilde{b})} {}_{(\tilde{c})} f_2^{(\tilde{c})}
\end{equation}
where $(\tilde{a})$ denotes an index string $a_1 a_2 \dots a_{r'}$ and
${\eta'}_{(\tilde{a})(\tilde{b})}$ is constructed as in
\eqref{metricdef} out of ${\eta'}_{ab}$, the induced metric on $V'$.
Since $\mathbb{V}'_1$ and $\mathbb{V}'_2$ have Young symmetrizers
whose tableaux are of differing shape, we can conclude (assuming
Theorem \ref{orthirreds} holds) that $F_1^{(a)} F_2^{(b)}
\eta_{(a)(b)} = 0$, and thus that $\mathbb{W}_1$ and $\mathbb{W}_2$
are orthogonal under the inner product $\eta_{(a)(b)}$.

To prove Theorem \ref{orthirreds}, we must define two new objects.
First, we define a linear map $\mathfrak{C}: \mathfrak{s}_r \to
\mathfrak{s}_r$ such that if $a$ is a basis element of
$\mathfrak{s}_r$ corresponding to a permutation $\sigma$ (i.e., $a$ is
of the form \eqref{etaperm}), then $\mathfrak{C}(a)$ is the basis
element corresponding to the permutation $\sigma^{-1}$.  Since
$\mathfrak{C}$ is a linear map, its action on the basis elements
\eqref{etaperm} then defines its action on the entire algebra
$\mathfrak{s}_r$.  For an arbitrary $a \in \mathfrak{s}_r$, we will
write $\hat{a} \equiv \mathfrak{C}(a)$.  It is not hard to see (from
distributivity and the properties of inverses of products) that
$\widehat{ab} = \hat{b}\hat{a}$.  More importantly, we also have the
identity
\begin{align}
  \eta_{(a_i)(b_i)} \eta^{(b_{\sigma(i)})} {}_{(c_i)} &=
  \eta_{(a_i)(b_i)} \eta^{(b_i)} {}_{(c_{\sigma^{-1}(i)})} \notag\\
  &= \eta_{(a_i)(c_{\sigma^{-1}(i)})} \notag \\
  &= \eta_{(c_{\sigma^{-1}(i)})(a_i)} \notag \\
  & = \eta_{(c_i)(b_i)} \eta^{(b_{\sigma^{-1}(i)})} {}_{(a_i)}.
\end{align}
This implies that for an arbitrary element $A \in \mathfrak{s}_r$, we
have
\begin{equation}
\eta_{(a_i)(b_i)} A^{(b_i)} {}_{(c_i)} = \eta_{(c_i)(b_i)}
\hat{A}^{(b_i)} {}_{(a_i)}.
\end{equation}

Second, given a Young symmetrizer $Y\in \mathfrak{s}_r$ there exists an element
$\epsilon \in \mathfrak{s}_r$, defined by
\begin{equation}
  \label{epsdef}
  \epsilon = \frac{1}{\mu} \sum_{t \in S_r} t Y \hat{t}
\end{equation}
where the sum runs over all basis elements of the form \eqref{etaperm}
and $1/\mu$ is the coefficient (in $Y$) of the basis element
corresponding to the identity permutation $e$.  (For example, in
\eqref{symex} and \eqref{asymex}, $\mu = 2$.)  It can be shown that
$\epsilon$ possesses the following properties:\footnote{See Weyl
  \cite{Weyl}, particularly \S IV.3, for proof; note that our ``Young
  symmetrizers'' are the ``primitive idempotents'' discussed there.
  The only one of these properties which is not explicitly stated by
  Weyl is \eqref{epsselfconj}; it follows, however, from the fact that
  $\epsilon$'s components in terms of the basis \eqref{etaperm} are
  equal for all elements in the same conjugacy class of the underlying
  group, and that for the group $S_r$, all elements are in the same
  conjugacy group as their inverses.}
\begin{equation}
  \epsilon A = A \epsilon \quad \forall \quad A \in \mathfrak{s}_r,
\end{equation}
\begin{equation}
  \epsilon Y = Y,
\end{equation}
and
\begin{equation}
  \hat{\epsilon} = \epsilon.
  \label{epsselfconj}
\end{equation}
Moreover, if $\epsilon_1$ is derived (via \eqref{epsdef}) from a Young
symmetrizer $Y_1$, and $\epsilon_2$ is derived from a Young
symmetrizer $Y_2$, then we have
\begin{equation}
  \epsilon_1 \epsilon_2 = \epsilon_1 = \epsilon_2
\end{equation}
if $Y_1$ and $Y_2$ are derived from Young tableaux of the same shape,
and
\begin{equation}
  \epsilon_1 \epsilon_2 = 0
\end{equation}
otherwise.  We can now provide a simple proof of Theorem
\ref{orthirreds}:
\begin{proof} Since $F_1^{(a)} \in \mathbb{V}_1$, $Y_1^{(a)} {}_{(b)}
  F_1^{(b)} = F_1^{(a)}$.  Similarly, $Y_2^{(a)} {}_{(b)}
  F_2^{(b)} = F_2^{(a)}$.  Thus, we have
\begin{align*}
  \eta_{(a)(b)} F_1^{(a)} &P^{(b)} {}_{(c)} F_2^{(c)} \\ 
  &= \eta_{(a)(b)}
  Y_1^{(a)} {}_{(d)} F_1^{(d)} P^{(b)} {}_{(c)} Y_2^{(c)} {}_{(e)}
  F_2^{(e)} \\
  &= \eta_{(a)(d)}
  \hat{Y}_1^{(a)} {}_{(b)} (P Y_2)^{(b)} {}_{(e)}  F_1^{(d)} F_2^{(e)} \\
  &= \eta_{(a)(d)} (\hat{Y}_1 P Y_2)^{(a)} {}_{(e)} F_1^{(d)} F_2^{(e)}.
\end{align*}
But by the above properties of the $\epsilon$'s, $\hat{Y}_1 P Y_2 =
\widehat{\epsilon_1 Y_1} P \epsilon_2 Y_2 = \hat{Y}_1 \epsilon_1
\epsilon_2 P Y_2 = 0$, which vanishes because (by hypothesis) $Y_1$
and $Y_2$ have differing tableau shape.  Thus,
\begin{equation*}
\eta_{(a)(b)} F_1^{(a)} P^{(b)} {}_{(c)} F_2^{(c)} = 0
\end{equation*}
as desired.\footnote{Note that this proof relies upon the fact that
  the tableaux for $\mathbb{W}_1$ and $\mathbb{W}_2$ have differing
  shape.  In fact, two subspaces of $\mathbb{U}_r$ with the same
  tableau shape but different Young symmetrizers (e.g., {\scriptsize
    \young(\ai\aii,\aiii)} and {\scriptsize \young(\ai\aiii,\aii)})
  will in general not be orthogonal under $\eta_{(a)(b)}$.}
\end{proof}

Finally, we can address the issue of the overall signature of
$\mathbb{V}$ under the metric $\eta_{(a)(b)}$.  As noted above
\eqref{gld1decomp}, under the subgroup $GL(d-1) \subset GL(d)$ of
purely spatial transformations, $\mathbb{V}$ decomposes into the
direct sum of several subspaces $\mathbb{W}_i$.  As in the case of
our decomposition of $\mathbb{U}_r$ into subspaces
$\bar{\mathbb{U}}_i$ of definite sign, the metric $\eta_{(a)(b)}$
(when restricted to each $\mathbb{W}_i$) will be non-degenerate and of
definite sign.  Specifically, if the Young tableau corresponding to
$\mathbb{W}_i$ is obtained from the tableau corresponding to
$\mathbb{V}$ by removing $p$ boxes from it, then the norm of any
tensor $F^{(a)} \in \mathbb{W}_i$ will be of the form
\begin{multline}
  \label{wietanorm}
  F^{(a)} F^{(b)} \eta_{(a)(b)} \\ = (\eta_{dd})^p \sum \eta_{i_1 j_1}
  \dots \eta_{i_r j_r} F^{i_1 \cdots d \cdots d \cdots i_r} F^{j_1
    \cdots d \cdots d \cdots j_r},
\end{multline}
where the summation here is over all possible placements of $2p$ $d$'s
into the index string of $F^{(a)}$, with the remaining indices $\{
i_1, \dots, i_r \}$ and $\{ j_1, \dots, j_r \}$ taking on values
between $1$ and $d-1$.  We can see from this equation that
$\eta_{(a)(b)}$ is either positive definite or negative definite on
each space $\mathbb{W}_i$: the summation in \eqref{wietanorm} is
clearly positive, and the sign of $\eta_{(a)(b)}$ on $\mathbb{W}_i$ is
then equal to $(\eta_{dd})^p = (-1)^p$.  Moreover, the subspaces
$\mathbb{W}_i$ span $\mathbb{V}$, and we have shown above that they
are orthogonal to one another; hence, we can find an orthonormal basis
for $\mathbb{V}$ by taking the union of the orthonormal bases for each
$\mathbb{W}_i$.\footnote{Incidentally, this shows that $\mathbb{V}$
  itself is non-degenerate under the metric $\eta_{(a)(b)}$ (as
  asserted in Section \ref{vacmansec}), since we have constructed an
  orthonormal basis for it.}  The signature $(n_+, n_-)$ of
$\mathbb{V}$ can thus be obtained by adding up the dimensionalities of
the $\mathbb{W}_i$ subspaces in two categories:
\begin{subequations}
\label{glsig}
\begin{equation}
  n_+ = \sum_{\text{even } \mathbb{W}_i} D(\mathbb{W}_i)
\end{equation}
and
\begin{equation}
  n_- = \sum_{\text{odd } \mathbb{W}_i} D(\mathbb{W}_i),
\end{equation}
\end{subequations}
where $D(\mathbb{W}_i)$ is the dimension of the subspace
$\mathbb{W}_i$, and a space $\mathbb{W}_i$ is ``even'' or ``odd'' if
its Young tableau is obtained from that of $\mathbb{V}$ by removing an
even or odd number of boxes, respectively.  Using this technique,
along with the usual rules for obtaining the dimensionality of
irreducible representations of $GL(d)$ and $GL(d-1)$ \cite{Hamermesh},
we obtain the second (``general'') column of Table \ref{sigtable}.

\subsubsection{Trace-free symmetrized tensor spaces}

In the previous subsection, we ascertained the signature of a
$GL(d)$-irreducible subspace $\mathbb{V} \subset \mathbb{U}_r$ by
decomposing it into orthogonal, non-degenerate subspaces of known
signature, and ``adding up'' the signatures of these subspaces
\eqref{glsig}. Of course, Lorentz symmetry is not invariance under the
entire group $GL(d)$, but rather invariance under the group $O(d-1,1)
\subset GL(d)$.  We now wish to determine the signatures of the
irreducible $O(d-1,1)$ spaces $\bar{\mathbb{W}}_i$ defined in
\eqref{sodecomp}.  In essence, our procedure to determine the
signatures of these spaces will turn out to be ``subtractive'' in the
same sense that our procedure for $\mathbb{V}$ was ``additive''.

To determine these signatures, we first prove a lemma:
\begin{lemma} In the $GL(d) \to O(d-1,1)$ decomposition of an
  invariant subspace $\mathbb{V}$ of rank $r \leq 5$ tensors into
  several $\bar{\mathbb{W}}_i$, the subspaces $\bar{\mathbb{W}}_i$ are
  orthogonal and non-degenerate.
\end{lemma}  
\begin{proof} Consider two tensors $F_1^{(a)} \in \bar{\mathbb{W}}_1$
  and $F_2^{(a)} \in \bar{\mathbb{W}}_2$.  Since these tensors are
  entirely contained in an invariant $O(d-1,1)$ subspace, only one
  term in the decomposition \eqref{tracefreedecomp} will be
  non-vanishing.  Without loss of generality, let the ranks of the
  $(f_{1i})$ and $(f_{2i})$ tensors in these terms be $r_1 \geq r_2$.
  We can see that if $r_1 \neq r_2$, the inner product of $F_1^{(a)}$
  and $F_2^{(a)}$ will necessarily involve taking the trace of the
  trace-free tensor $(f_{1i})$, and so will vanish.  If $r_1 = r_2$,
  then we might instead obtain some terms where the $r_1$ indices of
  $(f_{1i})$ and the $r_1$ indices of $(f_{2i})$ are all contracted
  (and possibly permuted.)  But since $r \leq 5$, the Ferrers diagrams
  corresponding to the spaces $\bar{\mathbb{W}}_1$ and
  $\bar{\mathbb{W}}_2$ are distinct; thus, by Theorem
  \ref{orthirreds}, these contractions will all vanish as well.  Thus,
  the spaces $\bar{\mathbb{W}}_i$ are all orthogonal to each other
  under $\eta_{(a)(b)}$.  Moreover, since $\mathbb{V}$ is
  non-degenerate under $\eta_{(a)(b)}$ and the orthogonal spaces
  $\bar{\mathbb{W}}_i$ span $\mathbb{V}$, it can be shown that each
  $\bar{\mathbb{W}}_i$ is non-degenerate under $\eta_{(a)(b)}$.
\end{proof}

Since the spaces $\bar{\mathbb{W}}_i$ are orthogonal and
non-degenerate under $\eta_{(a)(b)}$, we can see that the inner
product between two tensors will be of the form
\begin{equation}
  \eta_{(a)(b)} F_1^{(a)} F_2^{(b)} = \sum_i g_i(f_{1i}, f_{2i}),
\end{equation}
where the summation runs over the subspaces $\bar{\mathbb{W}}_i$, and
each $g_i$ is a non-degenerate quadratic form defined on
$\bar{\mathbb{W}}_i$.  More specifically, if $ F_1^{(a)}, F_2^{(a)}
\in \bar{\mathbb{W}}_i$, then only the term in \eqref{tracefreedecomp}
corresponding to this subspace will be non-vanishing; taking the inner
product of this term with itself, we have
\begin{multline}
\eta_{(a)(b)} F_1^{(a)} F_2^{(b)} = \eta_{(a)(b)} Y^{(a)} {}_{(c)} \left[ \eta^{c_1 c_2} \dots
  \eta^{c_{2t-1} c_{2t}} 
  f_{1i}^{(\tilde{c})} \right] \\ \times Y^{(b)} {}_{(d)} \left[ \eta^{d_1 d_2} \dots
  \eta^{d_{2t-1} d_{2t}}  f_{2i}^{(\tilde{d})} \right],
\label{nobrauer}
\end{multline}
where $r-2t$ is the rank of the tensor space corresponding to
$\bar{\mathbb{W}}_i$ and $(\tilde{a})$ denotes the index string $a_{2t+1}
\dots a_r$.  

Let us now imagine expanding out all the terms involving the metric
$\eta_{ab}$ and its inverse in \eqref{nobrauer}.  (This includes the
$Y^{(a)} {}_{(b)}$ terms, as they are simply sums of products of
$\eta^a {}_b$.)  Contracting the $\eta$'s together in each such term,
we can see that each term will fall into one of two categories:
\begin{itemize} 
\item When we contract the $\eta$'s with each other, we end up with a
  contraction of the form $\eta_{c_j c_k} f_{1i}^{(\tilde{c})}$ or
  $\eta_{c_j c_k} f_{2i}^{(\tilde{c})}$, where $j$ and $k$ are between
  $2t+1$ and $r$, inclusive; in other words, such terms will be
  proportional to the trace of $f_{1i}$ or $f_{2i}$.  Since these
  tensors are by definition trace-free, such terms will vanish.
\item When we contract the $\eta$'s with each other, we end up each
  index in $(\tilde{c})$ paired up with one in $(\tilde{d})$;  in
  other words, something proportional to 
  \begin{equation}
    \eta_{(\tilde{c})(\tilde{e})} \eta^{(\tilde{e}_{\sigma(i)})}
    {}_{(\tilde{d}_i)}.
  \end{equation}
  In the process of contracting the $\eta$'s with each other, we might
  also end up taking the trace of $\eta$ one or more times;  such
  traces will, of course, just give rise to factors of $d$.
\end{itemize}
Since the inner product of two tensors in $\bar{\mathbb{W}}_i$ will equal
the sum of contractions of this sort, we can then see that we must have
\begin{equation}
  g_i (f_{1i}, f_{2i}) = \eta_{(\tilde{a})(\tilde{b})}
  f_{1i}^{(\tilde{a})} P^{(\tilde{b})} {}_{(\tilde{c})}
  f_{2i}^{(\tilde{c})}
  \label{WiPdef}
\end{equation}
for some $P^{(\tilde{a})} {}_{(\tilde{b})} \in \mathfrak{s}_{r-2t}$.
The exact form of $P$ can in principle be determined (albeit
tediously) via the construction above.

I now make the following conjecture:
\begin{conj}
  Let $\mathbb{V}$ be an irreducible $GL(d)$ subspace of
  $\mathbb{U}_r$, and let $F_1^{(a)}, F_2^{(a)} \in \mathbb{V}$.  Then
  for any $P \in \mathfrak{s}_r$, there exists a number $\alpha$
  such that
  \begin{equation}
    \eta_{(a)(b)} F_1^{(a)} P^{(b)} {}_{(c)} F_2^{(c)} = \alpha
    \eta_{(a)(b)} F_1^{(a)} F_2^{(b)}. 
    \label{conjeq}
  \end{equation}
\end{conj}
I have been unable to prove this for general $r$;  however,
investigations via Mathematica have shown it to be true for the cases
of current interest (i.e., $r \leq 5$.)  If this conjecture holds,
then we will have
\begin{equation}
  \eta_{(a)(b)} F_1^{(a)} F_2^{(b)} = \sum_i \alpha_i
  \eta_{(\tilde{a})(\tilde{b})} f_{1i}^{(\tilde{a})}
  f_{2i}^{(\tilde{b})},
  \label{Winorm}
\end{equation}
where again the summation runs over the spaces $\bar{\mathbb{W}}_i$.  

Since the spaces $\bar{\mathbb{W}}_i$ are non-degenerate, we know that each
$\alpha_i \neq 0$; however, it is not immediately clear whether the
$\alpha_i$ coefficients are positive or negative.  To see that they
are in fact positive, suppose that we followed the above procedure for
the subgroup $O(d) \subset GL(d)$ rather than $O(d-1,1) \subset
GL(d)$; in other words, suppose that we were looking at the subset of
$GL(d)$ which left a positive definite quadratic form $\delta_{ab}$ on
$V$ unchanged.  The arguments leading up to \eqref{nobrauer} will
still hold (with $\eta$'s replaced with $\delta$'s).  Moreover, it is
not hard to see that the element $P$ of the algebra
$\mathfrak{s}_{r-2t}$ will be exactly the same for the case of $O(d)$
as the case of $O(d-1,1)$; the same terms will lead to the same traces
of trace-free tensors, the same permutations of the indices, and the
same traces of $\delta^{ab}$ (which are again equal to $d$).  Since
the conjecture only depends on the properties of the algebra
$\mathfrak{s}_r$ (and not the signature of the underlying space), the
coefficients $\alpha_i$ will thus be the same for the decompositions
$GL(d) \to O(d-1,1)$ and $GL(d) \to O(d)$.  But the coefficients
$\alpha_i$ must be positive in the case of $O(d)$:  if we take the
inner product of a tensor $F^{(a)}$ with itself under $\delta_{ab}$
rather than $\eta_{ab}$, we have
\begin{equation}
  \delta_{(a)(b)} F^{(a)} F^{(b)} = \alpha_i
  \delta_{(\tilde{a})(\tilde{b})} f_{i}^{(\tilde{a})}
  f_{i}^{(\tilde{b})}.
\end{equation}
The left-hand side of this equation is manifestly positive, while the
right-hand side must be of the same sign as $\alpha_i$.  Thus, the
coefficients $\alpha_i$ must be positive, both in the $O(d)$ inner
product and in the $O(d-1,1)$ inner product we are interested in.

With these results in hand, we can now finally answer the question of
the signature of the irreducible tensor subspaces of $O(d-1,1)$.
Suppose we construct an orthonormal basis for each of the subspaces
$\bar{\mathbb{W}}_i$, with respective signatures $(n_{i+}, n_{i-})$.  The
union of all of these bases will form a basis for $\mathbb{V}$.
Moreover, since all of the coefficients $\alpha_i$ in the
decomposition \eqref{Winorm} are positive, we can see that the
signature of each subspace $\bar{\mathbb{W}}_i$ under $\eta_{(a)(b)}$ will
be exactly the same as the signature (under
$\eta_{(\tilde{a})(\tilde{b})}$) of the lower-rank tensor space that
$\bar{\mathbb{W}}_i$ is similar to.  This then implies that if $\mathbb{V}$
has signature $(n_+, n_-)$, we will have
\begin{equation}
  (n_+, n_-) = \sum_i (n_{i+},n_{i-}),
\end{equation}
or, if $\bar{\mathbb{W}}_1$ is the subspace of $\mathbb{V}$ consisting of
all trace-free tensors of that same symmetry type,
\begin{equation}
  (n_{1+}, n_{1-}) = \left( n_+ - \sum_{i \neq 1}n_{i+}, n_- - \sum_{i
      \neq 1} n_{i-} \right).
\end{equation}

We can now see how, given knowledge of the signatures of all spaces of
trace-free tensors of rank less than $r$, to obtain the signature of
the spaces of all such tensors of a given symmetry type and rank $r$.
From the previous subsection, we know the signature $(n_+, n_-)$ of
$\mathbb{V}$.  We further know all of the signatures $(n_{i+},
n_{i-})$ for $i \neq 1$, since all the subspaces $\bar{\mathbb{W}}_i$
for $i \neq 1$ correspond to tensors of rank $r-2$ or lower.  Thus, we
can calculate $(n_{1+}, n_{1-})$ by taking the signature of
$\mathbb{V}$ and ``subtracting off'' the signatures of the other
subspaces (this is what we meant above by this procedure being
``subtractive''.)  We can then ``bootstrap'' our way up to higher and
higher tensor rank, starting with the signature of simple spaces like
the scalars ($r=0$) or vectors ($r=1$) and working our way up in rank,
calculating the signatures of the trace-free tensors of all symmetry
types for each rank.  The results of such a calculation are shown in
the third (``trace-free'') column of Table \ref{sigtable}.

\subsection{Discussion}

A few patterns are evident in the signatures shown in Table
\ref{sigtable}.  We note that the signatures for the signatures of
several of the trace-free tensor spaces are the same, but with $n_+$
and $n_-$ flipped; in particular, this is the case for tensors with
Ferrers diagrams $\{1 \}$ and $\{ 1,1,1 \}$; $\{2\}$ and $\{2,1,1\}$;
and $\{ 3 \}$ and $\{3,1,1\}$.  This is because these diagrams are
\emph{associates} of each other when $d=4$ \cite{Hamermesh}, and thus
these representations are equivalent under $O(3,1)$.  Since these
representations are equivalent, there must be a map between them; it
turns out to be the Hodge dual, obtained by contracting the volume
element $\epsilon^{abcd}$ with the antisymmetrized indices of the
first column in the appropriate Young symmetrizer.  The dual map will
take basis vectors to basis vectors; however, when we take their norm
in the new space, their sign will flip due to the identity
\begin{equation}
  \epsilon^{a_1 a_2 a_3 a_4} \epsilon_{b_1 b_2 b_3 b_4} = - 24
  \eta^{[a_1} {}_{b_1} \eta^{a_2} {}_{b_2} \eta^{a_3} {}_{b_3}
  \eta^{a_4]} {}_{b_4}.
\end{equation}
We can see that if we contract a vector $w^{(a)}$ in one space
$\bar{\mathbb{W}}$ with $\epsilon^{abcd}$, and then contract the
resulting basis vector in the associate space $\bar{\mathbb{W}}'$ with
itself, we will get a minus sign relative to what we would have
obtained had we simply taken the norm of $w^{(a)}$ in
$\bar{\mathbb{W}}$.  This also explains why $n_+ = n_-$ for any
representation whose Ferrers diagram consists of two rows (e.g.,
$\{1,1\}$, $\{2,1\}$, $\{3,1\}$, $\{2,2\}$, etc.): the spaces of
tensors of these symmetry types are mapped to themselves by the dual
map. Since positive-norm basis vectors are mapped to negative-norm
basis vectors and vice versa under the dual mapping, we must have $n_+
= n_-$ for these spaces.

The above classification system works for all tensors with rank $r
\leq 5$.  While tensors of higher rank than this are of rapidly
diminishing physical interest, it would be still be of interest to be
able to extend our discussion to such tensors.  Unfortunately, there
are two points in our procedure above that do not generalize
straightforwardly for tensors of rank $r \geq 6$: the lemma concerning
the orthogonality of the spaces $\bar{\mathbb{W}}_i$, and the
conjecture \eqref{conjeq}.  As the lemma stands, it relies upon the
fact that the invariant $O(d-1,1)$ subspaces of $\mathbb{V}$ are
uniquely labelled by the Ferrers diagrams of their corresponding
subspaces, and thus are orthogonal.  As noted in Footnote
\ref{rank6bad}, the decomposition of certain spaces $\mathbb{V}$ of
tensors with $r \geq 6$ will, in general, contain multiple subspaces
$\bar{\mathbb{W}}_i$ with the same underlying symmetry type; thus, we
cannot use Theorem \ref{orthirreds} to prove their orthogonality.  I
suspect that it is still true that these spaces are orthogonal, due to
properties of the Young symmetrizers; however, it is not immediately
obvious to me that this will be the case.

In the case of the conjecture \eqref{conjeq}, I am even more strongly
inclined to believe that this holds for all $r$.  In terms of the
algebra $\mathfrak{s}_r$, it is not too hard to see that this
conjecture is equivalent to the statement that $\hat{Y} P Y = \alpha
\hat{Y} Y$ for any Young symmetrizer $Y$ and any $P \in
\mathfrak{s}_r$;  in fact, the Mathematica calculations mentioned were
done by explicitly obtaining $\alpha$ for all $r!$ basis vectors in
$\mathfrak{s}_r$ and all Ferrers diagrams with $r$ boxes.  These
calculations are in principle doable for higher rank, but given the
growth of both the number of basis vectors and Young patterns with
increasing $r$ ($r=4$ requires 120 calculations;  $r=5$ requires 840;
$r=6$ would require 7920) it would be better to prove this once and
for all rather than on a case-by-case basis.

\bibliographystyle{apsrev}
\bibliography{lortop}{}

\end{document}